\documentclass[twocolumn,superscriptaddress,aps,pra]{revtex4-2}
\usepackage{bm}
\usepackage{amssymb,amsmath}
\usepackage{comment}
\usepackage{times}
\usepackage{braket} 
\usepackage[dvips]{graphicx}
\usepackage{array}
\usepackage[colorlinks,linkcolor=blue, urlcolor=blue, anchorcolor=blue, citecolor=blue]{hyperref}

\usepackage{ulem,lipsum}

\begin{document} 

\title{Dynamical Geometry of the Haldane Model under a Quantum Quench}

\author{Liwei Qiu}
\affiliation{Zhejiang Institute of Modern Physics and Zhejiang Key Laboratory of Micro-nano Quantum Chips and Quantum Control, Zhejiang University, Hangzhou 310027, China}

\author{Lih-King Lim}
\email{lihking@zju.edu.cn}
\affiliation{Zhejiang Institute of Modern Physics and Zhejiang Key Laboratory of Micro-nano Quantum Chips and Quantum Control, Zhejiang University, Hangzhou 310027, China}

\author{Xin Wan}
\email{xinwan@zju.edu.cn}
\affiliation{Zhejiang Institute of Modern Physics and Zhejiang Key Laboratory of Micro-nano Quantum Chips and Quantum Control, Zhejiang University, Hangzhou 310027, China}
\affiliation{CAS Center for Excellence in Topological Quantum Computation, University of Chinese Academy of Sciences, Beijing 100190, China}

\begin{abstract}
We explore the time evolution of a topological system when the system undergoes a sudden quantum quench within the same nontrivial phase.
Using Haldane's honeycomb model as an example, we show that equilibrium states in a topological phase can be distinguished by geometrical features, such as the characteristic momentum at which the half-occupied edge modes cross, the associated edge-mode velocity, and the winding vector about which the normalized pseudospin magnetic field winds along a great circle on the Bloch sphere.  
We generalize these geometrical quantities for non-equilibrium states and use them to visualize the 
quench dynamics of the topological system. 
In general, we find the pre-quench equilibrium state relaxes to the post-quench equilibrium state in an oscillatory fashion, whose amplitude decay as $t^{1/2}$. 
In the process, however, the characteristic winding vector of the non-equilibrium system
can evolve to regimes that are not reachable with equilibrium states. 
\end{abstract}

\date{\today}

\maketitle

%%%%%%%%%%%%%%%%%%%%%%%%%%%%%%%%%%%%%%%%%%%%%%%%%%%%%%%%%%%%%%%%%%%%%%%%%%%%%%%%%%%%%%%%%%%%%%%%%%%%
\section{Introduction}

Quantum entanglement measures nonclassical correlation among different parts of a quantum system 
and is an essential resource for quantum information and quantum computation~\cite{NielsenChuang}.
Of different entanglement properties, entanglement spectrum is particularly insightful as a diagnostic of
topological order, as first emphasized by Li and Haldane~\cite{Li08}.
Beyond fractional quantum Hall states~\cite{Sterdyniak12,Dubail12,Liu13,Rodriguez13},
it has been subsequently applied to a variety of topological systems, among which are 
topological band insulators~\cite{Fidkowski10,Turner10},
spin chains~\cite{Pollmann10,Thomale10},
Chern insulators~\cite{Prodan10,Huang12a, Huang12b,Hermanns14}, and 
Kitaev's honeycomb model~\cite{Yao10}.
Rapid development of quantum computers also makes it a reality to experimentally measure the entanglement spectrum of topological phases of matter~\cite{Choo18, Kokail21}. 

For topological quantum states, general correspondence exists between entanglement spectrum 
and the edge state spectrum~\cite{Qi12,Swingle12,Chandran11,Cirac11,Huang12a}.
In noninteracting fermion systems, the entanglement spectrum can be calculated efficiently 
through the spectrum of the correlation matrix in a subsystem~\cite{Peschel03}.
In an explicit calculation of Haldane's honeycomb lattice model, Huang and Arovas~\cite{Huang12a, Huang12b} 
identified the momentum $k_c$ of the half entanglement occupation where 
the zigzag edge state crossing and the half-odd-integer Wannier centers occur. 
The dynamics of entanglement spectrum after a quench has also been studied to detect the dynamical topology~\cite{Gong18,Chang18,Lu19}. 
In particular, the spectral crossings at half entanglement occupation has been proposed 
to be a robust topological signature~\cite{Gong18}. 
The crossings in the entanglement spectrum evolution can also reveal 
non-Hermitian dynamical topology of open and driven quantum systems~\cite{Sayyad21,Hermanns22}.

Due to the immense interest in topology in condensed matter physics, the studies of quench dynamics have focused on identifying the topological signatures in the time evolution of quantum many-body systems \cite{Unal16, Wang17, Tarnowski19, Sun18}.
The information beyond the topological characterization can also be of interest, 
not just for qualitative studies. 
In the fractional quantum Hall effect, Haldane~\cite{Haldane11} proposed that the Laughlin state 
should be understood as a family of wave functions~\cite{Qiu12}, all with the same topology, 
but different geometry. 
Geometrical parameter can thus be introduced for anisotropic quantum Hall systems to characterize the shape of correlation holes.
Quantum quenches under a sudden tilt of magnetic field, which changes the geometrical parameter, reveals the oscillatory response of the long-wavelength limit of the Girvin-MacDonald-Platzman magnetoroton~\cite{Liu18}. Geometrical quenches in bilayer fractional quantum Hall states can also induce 
nonequilibrium dynamics related to the collective dipole mode~\cite{Liu21}. 

It is, then, an interesting question to ask whether the indicator of topology 
in the entanglement spectrum can exhibit dynamical geometrical features that are 
related to a sudden quench within the same topological phase. 
In fact, the dynamical properties, such as the crossing momentum $k_c(t)$, are quantitatively well-defined (in an initial time window) \cite{Ji22},
just as dynamical topology \cite{Gong18,Chang18,Lu19,Sayyad21,Hermanns22}, even though the time-evolving wave function  does not match any of the equilibrium ground states~\cite{Ji22} with the same topological index. In this sense, they are driven purely by mismatches in geometric characterization of the system.   

In this paper, we use Haldane's honeycomb lattice model as a concrete example to systematically 
study the time evolution of systems with nontrivial topology under sudden quantum quench that 
preserves the topology. 
We first focus on the condition for the momentum $k_c$ of the half entanglement occupation, 
whose existence signals the presence of nontrivial topology.
We show that, using the pseudospin magnetic field formalism, we can map the dynamical problem 
into an equilibrium one, leading to a generalized equation for $k_c$ 
with a geometrical interpretation. 
On the one hand, a general geometrical quench leads to oscillation of $k_c$ with amplitude
decreases as $t^{-1/2}$.
In the stationary-phase approximation, the dominant frequencies and the amplitude decrease 
can be attributed to the coherent superposition of interband excitations. 
On the other hand, for pre- and post-quench equilibrium states with the same $k_c$, 
time evolution of the geometrical information is encoded in the oscillating slope of the edge mode, 
or edge-mode velocity, in the entanglement spectrum, whose amplitude increases as $t^{1/2}$. 
We unify the two apparently different quench processes with a unit-vector representation of the non-equilibrium system,
by the so-called winding vector \cite{Montambaux18, Lim20},  
and show that the time evolution of the topological system 
upon quench can be visualized by the winding vector trajectory on the Bloch sphere. 
As a result, we find that even quench within the same topological phase 
leads to states whose winding vector cannot be mapped to the equilibrium states. 

The paper is organized as follows. In Sec.~\ref{sec:model} we review Haldane's model and 
the solution for $k_c$ in the pseudospin magnetic field formalism that can be straightforwardly 
generalized to the non-equilibrium case. 
We derive the dynamical condition for the half-entanglement-occupation mode 
in Sec.~\ref{sec:noneq} and analyze the time evolution of $k_c$ 
in the stationary-phase approximation. 
In Sec.~\ref{sec:geometry} we discuss the additional information on the time evolution 
of the non-equilibrium system in the entanglement velocity.
We define the dynamical winding vector in Sec.~\ref{sec:winding} and show that its evolution 
can be used in generic cases to visualize the geometrical information in the non-equilibrium states. 
We summarize and discuss its potential generalizations in Sec.~\ref{sec:summary}.

%%%%%%%%%%%%%%%%%%%%%%%%%%%%%%%%%%%%%%%%%%%%%%%%%%%%%%%%%%%%%%%%%%%%%%%%%%%%%%%%%%%%%%%%%%%%%%%%%%%%
\section{Model and Equilibrium Properties}
\label{sec:model}

\subsection{Haldane's honeycomb model}
\begin{figure}
    \includegraphics[width=8cm]{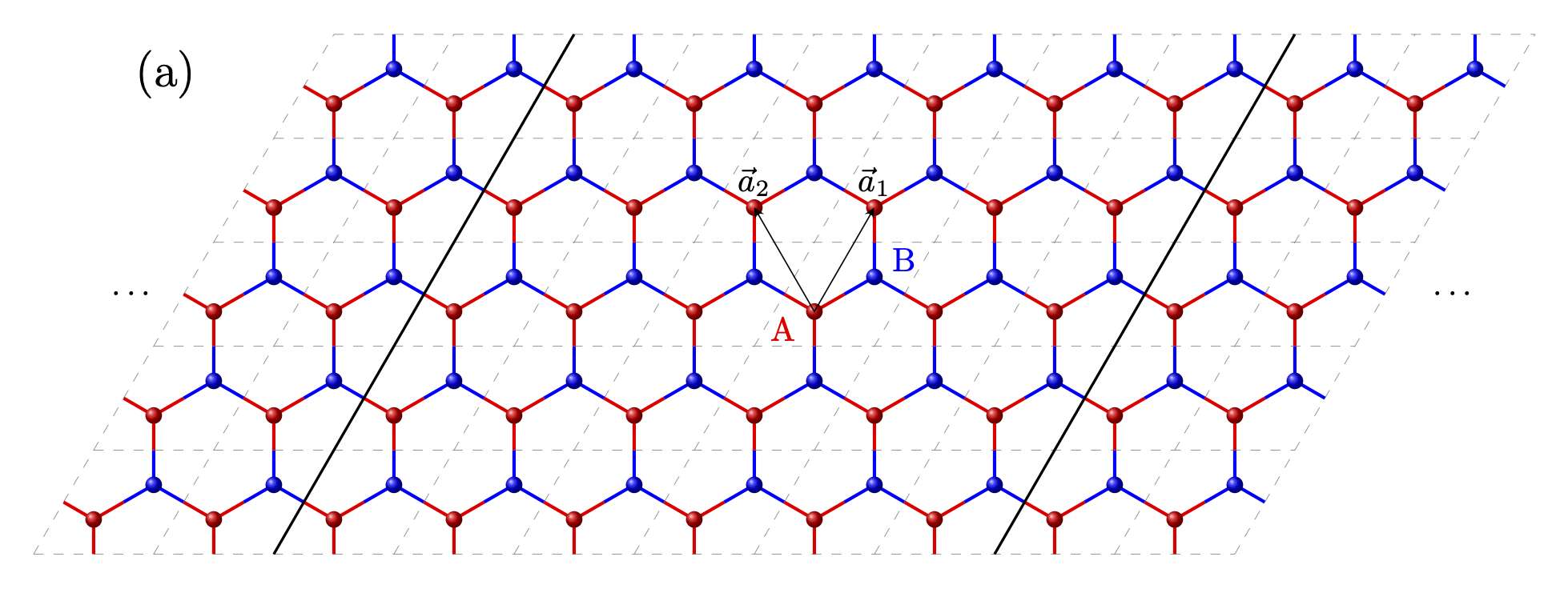}
    \includegraphics[width=7cm]{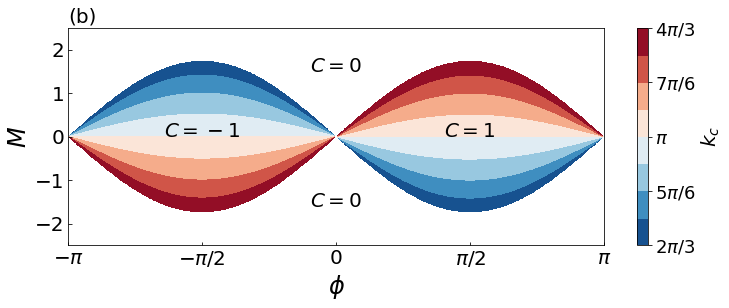}
    \caption{\label{fig:model}
    (a) 2D honeycomb lattice for the Haldane model, with the primitive vectors $\vec{a}_{1,2}=a (\pm \sqrt{3}/2,3/2)$ and the reciprocal vectors $\vec{a}_{1,2}^*=(4\pi/3a) (\pm \sqrt{3}/2,1/2)$ such that $\vec{a}^*_\mu\cdot\vec{a}_\nu=2\pi\delta_{\mu\nu}$. The $A$- and $B$-sublattice are related by a constant shift of the vector $a(0,-1)$. We set the lattice constant $a$ to unity. To calculate the OPES, we partition the system along $\vec{a}_1$ (thick black lines) into two subsystems with zigzag edges. 
    (b) The phase diagram of the model with topological phases ($C=\pm1$) surrounded by non-topological phases ($C=0$). The contour plot shows the crossing momentum $k_c$ for the zero modes. }
\end{figure}

To model a quantum Hall system without Landau levels, Haldane introduced a 2D honeycomb model for spinless lattice fermions with zero total flux~\cite{Haldane88}:
\begin{align}
    \label{eq_Hamiltonian}
        \hat{H} & =-t_0 \sum_{\langle i,j \rangle} \left(\hat{c}_i^{\dagger}\hat{c}_j +{\rm{h.c.}} \right) -t_1 \sum_{\langle \langle i,j \rangle \rangle} \left( e^{{\rm i}\phi_{ij}} \hat{c}_i^{\dagger}\hat{c}_j +{\rm{h.c.}} \right) 
        \nonumber \\
        & +M \left ( \sum_{i\in A}\hat{n}_i- \sum_{i\in B}\hat{n}_i \right ),
\end{align}
where $\hat{c}_i$ ($\hat{c}_i^\dag$) annihilates (creates) a fermion at site $i$, see Fig.~\ref{fig:model}(a). 
The nearest neighbor hopping strength is set to $t_0=1$, while the next-nearest neighbor hopping
strength $t_1= 1/3$ with time-reversal symmetry breaking phase $\phi_{ij} = \pm \phi$, whose sign depends on its direction. 
The inversion symmetry breaking on-site energy $M$ and $-M$ are introduced on $A$ and $B$ sites, respectively. 

Using periodic boundary conditions for the 2D lattice with $L$ unit cells, the operator $\hat{c}_i$ acting on position $i=(m,n)=m\vec{a}_1+n\vec{a}_2$, $m,n$ are integers, of the sublattice $\sigma=A,B$ in the momentum space is
\begin{align}
\hat{c}_{i}&=\frac{1}{\sqrt{L}}\,\sum_{k_1, k_2}\,e^{i(\frac{k_1}{2\pi}\vec{a}_1^*+\frac{k_2}{2\pi}\vec{a}_2^*)\cdot\,(m \vec{a}_1+n \vec{a}_2)}\,\hat{c}_{\sigma}(k_1,k_2),
\end{align}
with the primitive vectors $\vec{a}_{1,2}$, the reciprocal vectors $\vec{a}^*_{1,2}$, and the crystal momenta $k_{1,2}$ along $\vec{a}_{1,2}$ directions. The Bloch Hamiltonian can be written in a pseudospin form as 
\begin{equation}
H(\mathbf{k})=W(\mathbf{k})+ \vec{B}(\mathbf{k}) \cdot \vec{\sigma}
\end{equation} 
with $\mathbf{k} \equiv (k_1, k_2) = \frac{k_1}{2\pi}\vec{a}_1^*+\frac{k_2}{2\pi}\vec{a}_2^*$ and Pauli matrices $\vec{\sigma}=(\sigma_x,\sigma_y,\sigma_z)$. 
The pseudospin magnetic field $\vec{B}$ and the energy shift $W$ are
\begin{equation}
    \label{eq_B}
    \begin{cases}
        B_x = -1 - \cos k_1 - \cos k_2, \\
        B_y = \sin k_1+ \sin k_2, \\
        B_z = M+\frac{2}{3} \sin \phi [\sin k_1- \sin k_2 + \sin(k_2 - k_1)], \\
        W = -\frac{2}{3} \cos \phi[\cos k_1 + \cos k_2 + \cos(k_2 - k_1)].
\end{cases}
\end{equation}
The energies of the two bands are $E_{\pm}(\mathbf{k}) = W (\mathbf{k}) \pm \vert {\vec B} (\mathbf{k}) \vert$, and we consider a system with a fully filled lower band (annihilated by $\hat d_{-}(\mathbf{k})$ with $E_-(\mathbf{k})$) and an empty upper band (annihilated by $\hat d_{+}(\mathbf{k})$ with $E_+(\mathbf{k})$) 
with 
\begin{equation}
\begin{pmatrix}
\hat d_{+}(\mathbf{k})\\
\hat d_{-}(\mathbf{k})
\end{pmatrix}=\begin{pmatrix}\cos(\theta_\mathbf{k}/2)& \sin(\theta_\mathbf{k}/2)e^{-i\varphi_\mathbf{k}}\\\sin(\theta_\mathbf{k}/2)& -\cos(\theta_\mathbf{k}/2)e^{-i\varphi_\mathbf{k}}\end{pmatrix}
\begin{pmatrix}
\hat c_{A}(\mathbf{k})\\
\hat c_{B}(\mathbf{k})
\end{pmatrix},
\end{equation}
where $(\theta_\mathbf{k},\varphi_\mathbf{k})$ are the polar and azimuthal angles parameterizing $\hat{B}(\mathbf{k}) = \vec{B}(\mathbf{k})/\vert \vec B(\mathbf{k}) \vert $. The ground state wavefunction is
\begin{equation}
\vert {\Psi_{\rm gs}}\rangle=\left ( \prod_{\mathbf{k}\in \rm BZ} \hat d_{-}^{\,\,\,\dag}(\mathbf{k}) \right )\vert 0\rangle.
\end{equation} 
The Chern index of the ground state is $C=\frac{1}{2}[{\rm{sgn}}(M+\sqrt{3}\sin \phi)-{\rm{sgn}}(M-\sqrt{3}\sin \phi)]$. 
The phase diagram contains two topological phases ($C=\pm1$) and a trivial phase ($C=0$), as illustrated in Fig.~\ref{fig:model}(b). 
The phase boundary $M = \pm \sqrt{3} \sin \phi$ can be obtained by the gap vanishing equation $\vert {\vec B} (\mathbf{k}) \vert = 0$.

\subsection{Entanglement spectrum}
The system is set up on a torus periodic in both $\vec{a}_{1,2}$ directions, and we choose the subsystem $\alpha$ with two edges running in parallel to the $\vec{a}_{1}$ direction separated by distance $L_\alpha$, as shown in Fig.~\ref{fig:model}(a). The entanglement spectrum in such a free-fermion system can be obtained 
from the correlation matrix $\mathbb{C}$ restricted in the subsystem with matrix element~\cite{Chung01,Peschel03}
\begin{equation}
\label{cmatrix}
\mathbb{C}_{ij}=\langle {\Psi_{\rm gs}} \vert \hat c_i \hat c_j^{\dagger} \vert {\Psi_{\rm gs}}\rangle,
\end{equation}
where $i,j$ label two sites in the subsystem. With a half-filling ground state, we have
\begin{equation}
\mathbb{C}_{ij}=\frac{1}{L}\sum_{k_1,k_2}^{BZ}e^{i [(m-m')k_1+(n-n')k_2]}\,C(k_1,k_2) 
\end{equation}
where $i=(m,n)$, $j=(m',n')$ and 
\begin{eqnarray}
        C(k_1,k_2) &\equiv& 
\begin{pmatrix}
\langle \hat c_{A}(\mathbf{k}) \hat c_{A}^{\dagger}(\mathbf{k})\rangle &   \langle \hat c_{A}(\mathbf{k}) \hat c_{B}^{\dagger}(\mathbf{k})\rangle \\
\langle \hat c_{B}(\mathbf{k}) \hat c_{A}^{\dagger}(\mathbf{k})\rangle &   \langle \hat c_{B}(\mathbf{k}) \hat c_{B}^{\dagger}(\mathbf{k})\rangle
\end{pmatrix}\nonumber\\
        &=& \frac{\mathbb{I}}{2} + \frac{ \hat{B}(\mathbf{k}) \cdot \vec{\sigma}}{2},
\end{eqnarray}
where $\mathbb{I}$ is the $2\times 2$ identity matrix. Because of translation invariance of the groundstate, $\mathbb{C}_{ij}=\mathbb{C}_{i-j}$ and has a block-Toeplitz matrix structure.

Because the partition maintains the translation invariance along $\vec{a}_1$ direction, the full entanglement spectrum is organized into distinct sectors $k_1\in(0,2\pi]$ for
the correlation matrix $\widetilde{\mathbb{C}}(k_1)$ with elements
\begin{equation}
 \begin{aligned}
\widetilde{\mathbb{C}}_{nn'}(k_1) &= \frac{1}{L_2}\sum_{k_2=0}^{2\pi} e^{i(n-n')k_2} \,C(k_1,k_2)
 \end{aligned}
\label{cmn}
\end{equation}
where $n,n' = 1,\ldots,L_\alpha$ labels the unit cell along the $\vec{a}_2$ direction of the subsystem ($L_2$ is the total system length in that direction). The eigenvalues $\lambda(k_1)$ of the correlation matrix $\widetilde{\mathbb{C}}(k_1)$, 
known as the one-particle entanglement spectrum (OPES), 
are distributed between 0 and 1. Note that in the literature, e.g., Refs. \cite{Chung01,Peschel03}, the correlation matrix is conventionally defined as $\langle \hat{c}_j^{\dagger} c_i \rangle =\delta_{ij}- \mathbb{C}_{ij}$, where it shares the same eigenspectrum with $\mathbb{C}$ of Eq.~(\ref{cmatrix}).

An alternative way to understand is to define a quadratic entanglement Hamiltonian 
\begin{equation}
H_E = \sum_{a} \epsilon_{a} \hat{f}_{a}^{+}  \hat{f}_{a}
\end{equation} 
through $\lambda_{a} = 1/(e^{\epsilon_{a}} + 1)$, 
where $\hat f_{a}$ diagonalizes the correlation matrix defined in Eq.~(\ref{cmn}). 
The set of $\epsilon_{a}$ are commonly referred to as the entanglement spectrum, 
while the OPES can be thought of as its Fermi-Dirac distribution function 
at temperature $k_B T = 1$, or the entanglement occupation.

\begin{figure}
    \includegraphics[width=8cm]{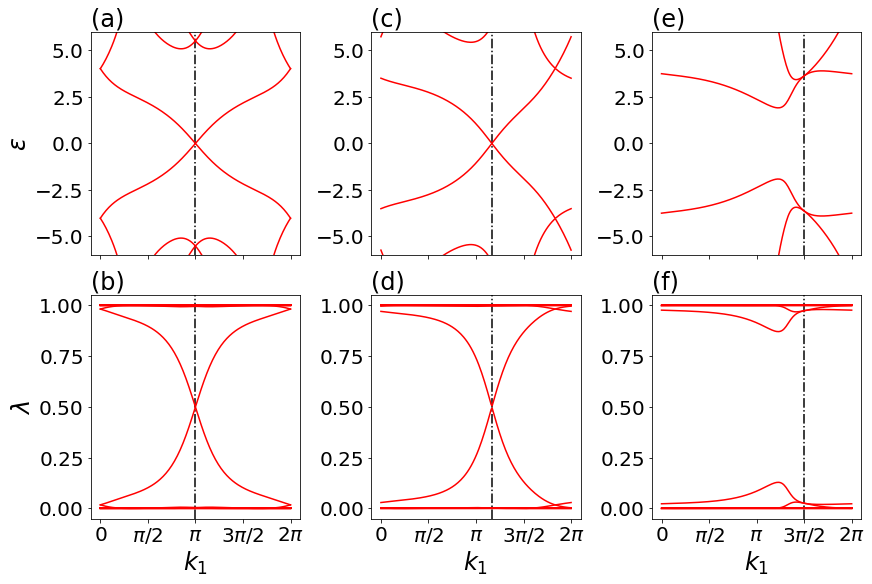}
    \caption{\label{fig:ES}
    Equilibrium entanglement spectra (a), (c), (e) and the corresponding OPES (b), (d), (f) 
    for the half-filled Haldane model with $L_2 = 100$, 
    which is partitioned into two subsystems $\alpha,\beta$ with width $L_{\alpha,\beta} = 50$. 
    The parameters are $M = 0$ and $\phi = \pi/2$ for (a) and (b), 
    $M = 1$ and $\phi = \pi/2$ for (c) and (d), 
    and $M = 2$ and $\phi = \pi/2$ for (e) and (f).   
    The system in a topological phase with Chern number $C = 1$ for (a)-(d), and the edge modes 
    in the spectra cross at $\epsilon = 0$ or $\lambda = 1/2$, indicated by the dot-dashed lines. 
    In the trivial phase ($M = 2$), the crossings in the spectra do not have zero modes. 
    }
\end{figure}

Examples of the entanglement spectra and OPES are shown in Fig.~\ref{fig:ES} 
for the three sets of $(M, \phi) = (0, \pi/2)$,  $(1, \pi/2)$, and $(2, \pi/2)$ for $L_{\alpha}=L_{\beta}= L_2/2=50$ (without loss of generality, we fix the ratio value $L_{\alpha,\beta}/L_2=1/2$ in the rest of the work). 
In the topologically nontrivial phase, the spectra feature counter-propagating edge modes 
that meet at a characteristic momentum $k_c$, at which $\epsilon = 0$ or 
$\lambda = 1/2$. 
We, thus, refer to the two degenerate modes at $\epsilon = 0$ as entanglement zero modes, 
which do not exist in topologically trivial phase, e.g., at $M = 2$ and $\phi = \pi/2$, 
as illustrated in Fig.~\ref{fig:ES}(e).  

As the existence of degenerate or quasi-degenerate in-gap edge modes in 1D systems, the existence of such entanglement zero modes are hallmark of the corresponding topological phase. 
But unlike in the 1D case, the degrees freedom in the extra dimension carries additional geometrical information that allows us to refine our characterization of the topological states. 

%%%%%%%%%%%%%%%%%%%%%%%%%%%%%%%%%%%%%%%%%%%%%%%%%%%%%%%%%%%%%%%%%%%%%%%%%%%%%%%%%%%%%%%%%%%%%%%%%%%%
\subsection{Equilibrium entanglement crossing}
\label{sec:eq_opes}

We now focus on the crossing in the OPES at $\lambda = 1/2$ or, equivalently, 
at $\epsilon = 0$ in the ES. 
For this purpose, we can restrict our discussion to the OPES, which can be simply obtained by 
diagonalizing the correlation matrix. 
When external parameters, such as $M$ and $\phi$, change within the same topological phase, 
the crossing or the pair of zero modes in the ES cannot be removed. 
However, the characteristic edge momentum $k_1 = k_c$, at which the zero modes appear, 
depends on the external parameters, as reflected in Figs.~\ref{fig:ES}(a) and (c).
They correspond to a left-going and a right-going mode on opposite sides of the subsystem. 

For the convenience of the later discussion, we highlight now the key steps 
in determining the location of $k_c$ in the context of OPES. We define a scaled-and-shifted correlation matrix 
$\mathbb{G} \equiv 2\,\mathbb{C} - \mathbb{I}$.
We note that in $\mathbf{k}$-space
\begin{equation}
G(\mathbf{k})= \hat{B}(\mathbf{k}) \cdot \vec{\sigma},
\end{equation}
which can be recognized as the operator 
that measures the pseudospin along the $\hat{B}(\mathbf{k})$ direction
and whose eigenvalues vary between -1 and 1. 
Obviously, the $G(\mathbf{k})$ operator shares the same eigenstates as the original Hamiltonian and 
can be used to identify phases. 

Similarly, for each $k_1$ sector, we have
\begin{equation}
\widetilde{\mathbb{C}}(k_1) = \frac{\mathbb{I}}{2} + \frac{\widetilde{\mathbb{G}}(k_1)}{2},
\label{gmatrix}
\end{equation}
where $\widetilde{\mathbb{G}}(k_1)$ is a block-Toeplitz matrix, 
in which all 2$\times$2 blocks in any diagonal are the same partial Fourier transform of $G(\mathbf{k})$, such that
\begin{equation}
\widetilde{\mathbb{G}}(k_1) = \begin{pmatrix}
g_{0} & g_{1} & \cdots & g_{L_{\alpha}-1} \\
g_{-1} & g_{0}   & \cdots &\vdots \\
\vdots & \vdots  & \ddots  &\vdots\\
g_{-(L_{\alpha}-1)}   & g_{-L_{\alpha}+2}   &\cdots  & g_{0}
\end{pmatrix},
\label{eq:toeplitz}
\end{equation}
where 
\begin{equation}
g_{n}(k_1) = \frac{1}{2\pi} \int_0^{2\pi} e^{ink_2} \,G(k_1, k_2) dk_2
\end{equation}
for $n=0,\ldots, L_{\alpha-1}$. Here we take $(1/L_2)\sum_{k_2}\rightarrow  (1/2\pi)\int dk_2$ for large $L_2$. Notice that the eigenvalues of $\widetilde{\mathbb{G}}(k_1)$ fall into the range of $[-1,1]$ and the spectral crossing corresponds to vanishing eigenvalues. 

Huang and Arovas~\cite{Huang12b} showed that in the equilibrium case the entanglement spectrum crossing, the edge spectrum crossing, and the half-odd-integer Wannier center coincide at the same $k_c$ and calculated explicitly with the following wave-function ansatz for edge modes:
\begin{equation}
\vert \psi \rangle = \vert \psi_A \rangle \otimes \ket{\beta}, 
\label{eq:ansatz}
\end{equation}
where $\vert \psi_A \rangle$ is the wave function
on sublattice $A$ describing its decay into the bulk, 
and 
\begin{equation}
\ket{\beta} = \frac1{1 + \vert \beta \vert^2} 
\begin{pmatrix}
1 \\ \beta
\end{pmatrix},
\end{equation}
where $\beta$ is a complex number, 
describes the relative wave function in the unit cell.
At the crossing where $\lambda = 1/2$, we have 
\begin{equation}
\label{bpsi}
\bra{\beta_c} \otimes \langle \psi_A \vert 
\widetilde{\mathbb{G}}(k_c) 
\vert \psi_A \rangle  \otimes \ket{\beta_c} = 0.
\end{equation}
We note that a sufficient condition to solve the matrix equation is to ignore 
the structure of $\vert \psi_A \rangle$ and demand
\begin{equation}
\bra{\beta_c}  g_{n}(k_c) \ket{\beta_c} = 0
\end{equation}
for all 2$\times$2 blocks of $\widetilde{\mathbb{G}}(k_c)$. We can further require
\begin{equation}
\label{bsigmabeta}
\bra{\beta_c} \hat B(k_c, k_2) \cdot \vec \sigma \ket{\beta_c} = 0
\end{equation}
for all $k_2$. We note that the operator
\begin{equation} 
\hat B(k_c, k_2) \cdot \vec \sigma = e^{i \pi \hat B(k_c, k_2) \cdot \vec \sigma / 2}
\end{equation}
is a unitary operator that rotates the state by an angle $\pi$ about the $\hat B(k_c, k_2)$ axis,
and Eq.~(\ref{bsigmabeta}) states that its expectation value vanishes, i.e.,
\begin{equation}
\langle \sigma \rangle_{c} \propto \bra{\beta_c} \hat B(k_c, k_2) \cdot \vec \sigma 
\ket{\beta_c} = 0.
\end{equation}
Geometrically, this means that $\hat B(k_c, k_2)$ are coplanar 
and lie on a great circle of the Bloch sphere.
If we define the normal direction of the circle as $\hat m(k_c)$,
the two eigenstates of $\hat m(k_c) \cdot \vec \sigma$ are $\vert {\beta_c}  \rangle$ and $[\hat B(k_c, k_2) \cdot \vec \sigma] \vert {\beta_c}  \rangle$ for any $k_2$. 
Alternatively, we can write the coplanarity condition of the pseudospin magnetic field $\vec{B}$
as
\begin{equation}
    \label{eq:coplanarity}
    \left[\hat{B}(k_c,k_2)\times\hat{B}(k_c,k_2^{\prime})\right]\cdot\hat{B}(k_c,k_2^{\prime\prime})=0
\end{equation}
for arbitrary $k_2$, $k'_2$, $k''_2$. 

For the Haldane model, the pseudospin magnetic field can be alternatively written as
\begin{equation}
\vec{B}(k_1, k_2) = \vec{n}_0(k_1) +  \vec{n}_1(k_1)  \cos{k_2} + \vec{n}_2(k_1)  \sin{k_2},
\label{equation:elliptical}
\end{equation} 
whose tip yields an ellipse for a given $k_1$. 
The center of the ellipse is $\vec{n}_0$, while the two axes are $\vec{n}_1$ and $ \vec{n}_2$.
They are explicitly given by 
\begin{equation}
        \begin{aligned}
            \vec{n}_0 &= \left (-1 - \cos k_1, \sin k_1, M + \frac{2}{3}\sin\phi\sin k_1 \right ),
            \\
            \vec{n}_1 &= \left (-1, 0, -\frac{2}{3}\sin{\phi} \sin{k_1} \right ),
            \\
            \vec{n}_2 &= \left (0, 1, \frac{2}{3}\sin{\phi} (\cos{k_1}-1) \right ).
        \end{aligned}
\end{equation}   
One can, equivalently, state the coplanarity condition as 
the coplanarity of $\vec{n}_{0}$, $\vec{n}_{1}$, and $\vec{n}_{2}$, i.e., 
\begin{equation}
\vec{n}_0 \cdot (\vec{n}_1 \times \vec{n}_2) = 0,
\label{eq:coplanarity2}
\end{equation}
and we can identify $\hat m(k_c)$ as the unit vector along $\vec{n}_1 \times \vec{n}_2$.
Therefore, we obtain the equation of $k_c$ as
\begin{equation}
\sin k_c=-M/(2 \sin\phi).
\label{equation:kc}
\end{equation} 
This, however, is not a sufficient condition, as $\cos k_c$ can be either positive or negative. 
For nontrivial topology to exist, the origin must be enclosed by the ellipse;
in other words, we must have 
\begin{equation}
\left (\frac{\vec{n}_0 \cdot \vec{n}_1}{\vec{n}_1 \cdot \vec{n}_1} \right )^2+ \left (\frac{\vec{n}_0 \cdot \vec{n}_2}{\vec{n}_2 \cdot \vec{n}_2} \right )^2 < 1,
\label{eq:topology}
\end{equation}
which rules out the solution for $\cos k_c > 0$ and $\vert \sin k_c \vert \geq \sqrt{3}/2$. 
In other words, $k_c$ is restricted between $2\pi/3$ and $4\pi/3$, specifically, 
between $2\pi/3$ and $\pi$ for $C = -1$ and between $\pi$ and $4\pi/3$ for $C = 1$. 
The contour plot of $k_c$ for various $M$ and $\phi$ is shown in Fig.~\ref{fig:model}(b) 
for the two nontrivial phases. 

\begin{figure}
    \includegraphics[width=8cm]{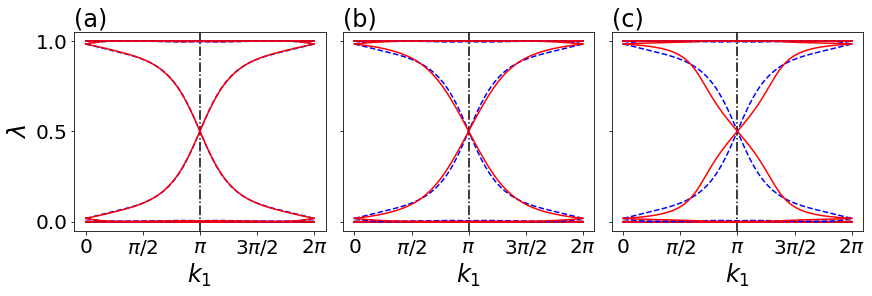}
    \includegraphics[width=8cm]{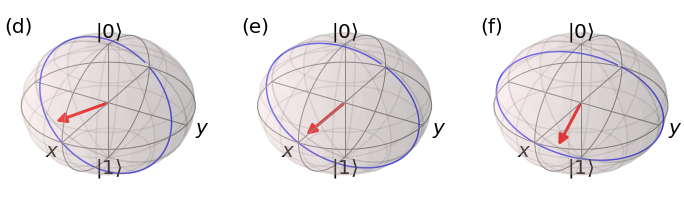}
    \caption{\label{fig:ES2}
    (a)-(c) OPES and (d)-(f) the corresponding normalized pseudospin magnetic field 
    $\hat B(k_c, k_2)$ at $k_c = \pi$ and its normal direction $\hat m (k_c)$ 
    for the half-filled Haldane model with $L_2 = 100$. 
    We fix $M = 0$ and choose $\phi = \pi/3$ for (a) and (d), 
    $\phi = \pi/6$ for (b) and (e), and $\phi = \pi/12$ for (c) and (f).   
    For comparison, we plot the OPES for $M = 0$ and $\phi = \pi/2$ by dashed lines in (a)-(c). }
\end{figure}
We note that even though $k_c$ can be used to characterize topological states 
with various combinations of $(M, \phi)$, the single number alone is not complete in distinguishing
all possible state. 
For example, all states along $M = 0$ share the same $k_c = \pi$, as illustrated in Fig.~\ref{fig:ES2}. 
The difference in the OPES lies at the slopes of the edge modes. 
An alternative way to visualize the difference is to plot the unit vector 
\begin{equation}
\hat m(k_c) = \frac{1}{2\pi} \int \hat B \times d\hat B,
\end{equation}
where the derivation and integral are carried out with respect to $k_2$ at fixed $k_1 = k_c$. 
Geometrically, $\hat{m}(k_c)$ is normal to the pseudospin magnetic field 
$\hat B(k_c, k_2)$, i.e., along 
\begin{equation}
\vec{n}_1 \times \vec{n}_2 = \left ( \frac{2}{3} \sin \phi \sin k_c, \frac{2}{3} \sin \phi (\cos k_c - 1), -1\right ),
\label{eq:mvec}
\end{equation}
as shown in Fig.~\ref{fig:ES2}(d)-(f).
Its dependence on $M$ is encoded in the expression of $k_c$ in Eq.~(\ref{equation:kc}). 
It is worthy pointing out that when the chiral symmetry is preserved, i.e., $M = 0$ or $\sin k_c = 0$, 
the vector $\hat{m}(k_c)$ is restricted in the $y$-$z$ plane. 
In particular, the point with $M = 0$ and $\phi = \pi/2$ at the center of the $C = 1$ phase has 
\begin{equation}
\hat m(k_c = \pi) = \left (0, -0.8, -0.6 \right ).
\end{equation}
While the change of $\phi$ leads to a rotation of $\hat m(k_c = \pi)$ in the $y$-$z$ plane,
a small change of $M$ by $\delta \ll 1$ leads to a linear change in $\sin k_c$, hence we find
\begin{equation}
\hat m(k_c \approx \pi + \delta/2) \approx \left (-0.2 \delta, -0.8, -0.6 \right ),
\end{equation}
i.e., a rotation of $\hat m(k_c = \pi)$ away from the $y$-$z$ plane. 
Therefore, we find that $\hat{m} (k_c)$ is a more complete description of the topology of the corresponding system than the identification of the zero modes at $k_1 = k_c$. 

\begin{figure}
    \includegraphics[width=4cm]{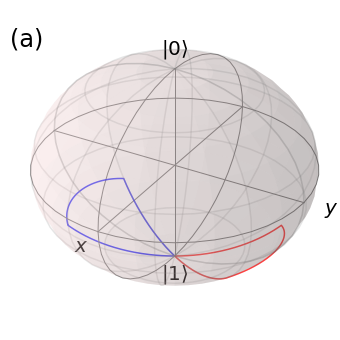}
    \includegraphics[width=4cm]{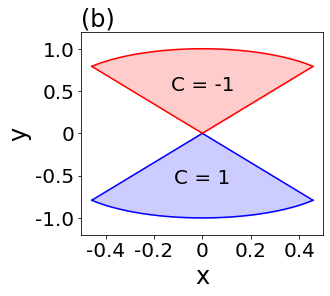}
    \caption{\label{fig:projection}
    (a) The regions covered by the head of $\hat m (k_c)$  on the Bloch sphere 
    for the two topological phases 
    with $C = 1$ (enclosed by blue curves) and $C = -1$ (enclosed by red curves). 
    (b) The stereographic projection of the two topological regions in (a). 
    }
    \end{figure}
The study of winding characteristics of $\hat{B}(\mathbf{k})$ appears in graphene ribbons relating to the presence/absence of edge modes \cite{Delplace11}. There, the quasi-one-dimensional problem is a graphene of finite extent along one direction. The presence of edge states (or zero modes) in the energy spectrum, depending on the types of termination, is equivalent to a non-trivial Zak phase for $\hat{B}(k_c, k_2)$ for $k_2$ running from $0$ to $2\pi$. More recently, the concept of supplementing the topological index by a unit vector with geometrical information 
also occurs in the context of merging and emergence of a pair of Dirac points~\cite{Montambaux18,Lim20}. 
One defines the winding vector with
\begin{equation}
\hat w = \frac{1}{2\pi} \int \hat n \times d\hat n,
\end{equation}
where $\hat n = \vec h / \vert \vec h \vert$ is the normalized psudospin magnetic field $\vec h$ 
for a Bloch Hamiltonian $\mathcal{H}(\mathbf{k}) =  \vec h(\mathbf{k}) \cdot \vec \sigma$. 
As in the present case, the inclusion of the geometry of winding axes is crucial 
in revealing the evolution of the pair of Dirac points.  
For the same reason, we refer to the unit vector $\hat m(k_c)$ as the winding vector, which provides further information than the momentum crossing $k_c$. 

Fig.~\ref{fig:projection}(a) illustrates the regions on the Bloch sphere that are covered
by the head of the winding vector $\hat m (k_c)$ for the two topological phases 
with $C = 1$ (enclosed by blue curves) and $C = -1$ (enclosed by red curves). 
The two regions head-to-head touch at state $\ket{1}$, 
which corresponds to the critical state with $M = 0$ and $\phi = 0$ where a topological phase transition 
from $C = 1$ to $C = - 1$ occurs. 
We plot the stereographic projection of the two regions to the plane tangent to state $\ket{1}$ in Fig.~\ref{fig:projection}(b). 
In this plot the two regions map to two fans connected at their vertices head-to-head. 
The straight lines map to the boundaries of the topological phases, i.e., $k_c = 2\pi /3$ and $4\pi/3$, 
or $M = \pm \sqrt{3} \sin \phi$. 
Interestingly, the two arcs map to the $\phi = \pm \pi/2$, which is inside the corresponding topological phases.
As $(M, \phi)$ and $(M, \pi - \phi)$ represent the same state, the pair $(M, \phi)$ form a two-to-one 
mapping to $\hat m (k_c)$. 
Therefore, we conclude that only a small portion of the Bloch sphere represents 
the equilibrium topological phases in the Haldane model.

%%%%%%%%%%%%%%%%%%%%%%%%%%%%%%%%%%%%%%%%%%%%%%%%%%%%%%%%%%%%%%%%%%%%%%%%%%%%%%%%%%%%%%%%%%%%%%%%%
\section{Non-equilibrium entanglement crossing}
\label{sec:noneq}

We now consider the dynamical case under a sudden change of the Hamiltonian 
from $H$ to $H^{\prime}$ at $t = 0$. 
The state of the system after the quantum quench evolves unitarily as $\vert \Psi(t)\rangle=e^{-{\rm i}H^{\prime}t}\vert \Psi_{\rm gs} \rangle$, where we set $\hbar=1$. 
The time-dependent correlation matrix in $k$-space, again, has a $2 \times 2$ form as 
\begin{equation}
    \label{eq_ckt}
    C(t, k_1, k_2) = \frac{\mathbb{I}}{2}+ \frac {\hat{B}(t, k_1, k_2) \cdot \vec{\sigma}}{2},
\end{equation}
where $\hat{B}(t, k_1, k_2)$ is the dynamical pseudospin magnetic field, which can be regarded as 
the pre-quench $\hat B$ precessing about the post-quench $\hat B'$ 
with a Larmor frequency $2 \vert \vec{B}' \vert$. Explicitly, we can express
$\hat{B}(t)$ as the sum of three mutually orthogonal vectors: 
\begin{equation}
    \label{eq_Bt}
    \hat{B}(t)=  \cos(2\vert \vec{B}' \vert t)  \vec{b}_1  + \sin(2\vert \vec{B}' \vert t) \vec{b}_2
    + \vec{b}_3.
\end{equation}
where 
\begin{align}
\vec{b}_1 &= \hat{B}-(\hat{B}\cdot \hat{B}^{\prime})\hat{B}^{\prime}, \label{eq:b1}\\
\vec{b}_2 &= \hat{B}\times \hat{B}^{\prime}, \label{eq:b2} \\
\vec{b}_3 &= (\hat{B}\cdot \hat{B}^{\prime})\hat{B}^{\prime}. \label{eq:b3}
\end{align}
Note that $\vec{b}_1 \cdot \vec{b}_1 = \vec{b}_2 \cdot \vec{b}_2 = 1 - \vec{b}_3 \cdot \vec{b}_3$. 
Therefore, the calculation of the dynamical OPES is technically identical to that in the equilibrium case. 
While the long-time quench behavior is determined by the time-independent $\vec{b}_3$,
we are interested here in the short-time dynamics determined by the oscillatory terms in the plane perpendicular to $\vec{B}'$. 

As in the equilibrium case, we can define 
$G(t, k_1, k_2)= \hat{B}(t, k_1, k_2) \cdot \vec{\sigma}$, 
so the time-dependent correlation matrix in the subsystem is 
\begin{equation}
\widetilde{\mathbb{C}}(t, k_1) = \frac{\mathbb{I}}{2} + \frac{\widetilde{\mathbb{G}}(t, k_1)}{2},
\label{gmatrix_t}
\end{equation}
where $\widetilde{\mathbb{G}}(t, k_1)$ has the identical form as in Eq.~(\ref{eq:toeplitz}).
In each $k_1$ block, we have
\begin{equation}
\label{eq:Gntk1}
g_{n}(t, k_1) = \frac{1}{2\pi} \int_0^{2\pi} e^{ink_2} G(t, k_1, k_2) dk_2.
\end{equation}

\begin{figure}
    \includegraphics[width=4cm]{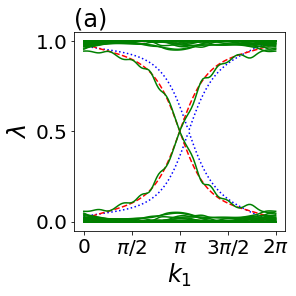}
    \includegraphics[width=4cm]{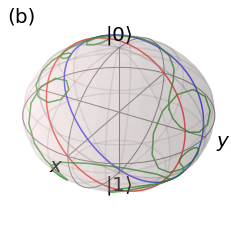}
    \caption{\label{fig:geometry}
    (a) Non-equilibrium OPES (green solid lines) at $t = 8$, after the system quenches from $M = \sqrt{3}/3$ to $M = 0$ at $t = 0$ with fixed $\phi = \pi/2$. The OPES for the pre- (blue dotted lines) and post-quench (red dashed lines) systems are included for comparison. 
    (b) Dynamical pseudospin magnetic field $\hat{B}(t, k_c)$ at the crossing momentum $k_c(t=8)$ (green) is no  longer coplanar as those for the pre- (blue) and post-quench (red) Hamiltonians.}
    \end{figure}

As we have discussed, we restrict ourselves to the quantum quench within the same topological phase ($C = 1$), such that the topology of the system remains unchanged upon the quench. 
As an example, we consider a quench from $M = \sqrt{3}/3$ to $M = 0$ at $t = 0$ with fixed $\phi = \pi/2$. In Figure~\ref{fig:geometry}(a) show the OPES for the dynamical system with $L_\alpha = L_2/2=50$ at $t = 8$.
Close to the equilibrium $k_c$ of the post-quench Hamiltonian, the spectrum has a well-defined zero-mode crossing, but the edge modes show wiggles away from the crossing. 
As demonstrated in Fig.~\ref{fig:geometry}(b), the coplanarity condition for $\hat B(t)$ is not valid in the dynamical case. 
In this section, we answer the following questions, which naturally arise from the observation above. 
How can one calculate the value of $k_c$ in the non-equilibrium case? 
What is the geometrical interpretation of the condition of the crossing? 
What can we learn from the evolution of $k_c$? 

Before we proceed to answer these questions, we emphasize the time scale of our interest.
According to the Calabrese-Cardy picture~\cite{Calabrese05,Calabrese06},
quasiparticles travel along $\vec{a}_2$ with a maximum velocity 
$v_2=\left . {\rm d}E/{\rm d} k_2 \right \vert_{\rm max}$ 
after the quench, 
which defines a characteristic time $t_2 = L_{\alpha} / v_2$ in our system.
For $0 < t < t_2$, quasiparticles excited from the zero mode at one edge of the subsystem $\alpha$ 
(which is assumed to be no larger than the other subsystem $\beta$)
do not have enough time to travel to the other edge, the degenerate zero modes do not have significant 
overlap,  so they do not split. 
During this time interval, the instantaneous entanglement crossing momentum $k_c(t)$ is, thus, 
well defined. 
As we restrict ourselves to $0 < t < t_2$, we effectively take the 
thermodynamic limit $L_{\alpha} \rightarrow \infty$ (hence $L_2 \rightarrow \infty$) 
before we take the long-time limit $t \rightarrow \infty$.

%%%%%%%%%%%%%%%%%%%%%%%%%%%%%%%%%%%%%%%%%%%%%%%%%%%%%%%%%%%%%%%%%%%%%%%%%%%%%%%%%%%%%%%%%%%%%%%%%
\subsection{Conditions for the non-equilibrium entanglement crossing}
\label{sec:noneq_cond}

In the non-equilibrium case, to follow the evolution of the geometric winding vector, we need to extract the evolution of an effective great circle traced by $\hat{B}(k_c,k_2)$ for $k_2$ running from $0$ to $2\pi$, from the pre-quench great circle at time $t=0$ to the post-quench ones at $0\ll t<t_2$ with a generalized coplanarity condition. The lowest-order nontrivial approximation is equivalent to truncating $\widetilde{\mathbb{G}}(t, k_c)$, the time-dependent version of Eq.~(\ref{eq:toeplitz}), into a tridiagonal one:
 \begin{equation}
            \begin{aligned}
                \widetilde{\mathbb{G}}
                    \approx \left(\begin{array}{lllllll}
                     g_{0} &  g_{1}  & 0  & \cdots & 0 & 0  \\
                     g_{-1} & g_{0} &  g_{1} & \cdots & 0 & 0  \\
                    0 &  g_{-1} & g_{0}&  \ddots & 0  & 0 \\
                    \vdots & 0 & \ddots & \ddots & \ddots  & \vdots \\
                    0 & \vdots & \ddots &  g_{-1} &  g_{0} &  g_{1} \\
                    0 & 0 & \cdots & 0 &  g_{-1} &  g_{0} 
                    \end{array}\right)
                        \end{aligned}.
                        \label{equation:C_gamma=2}
        \end{equation}
In other words, $k_c$ can be solved by retaining in the inverse Fourier transform $G(t, k_c, k_2)$ from Eq.~(\ref{eq:Gntk1}) its first harmonics
\begin{equation}
\label{eq:truncation}
G_{\rm eff}(t, k_c, k_2)
\approx g_{0}(t, k_c) + [g_{1}(t, k_c)e^{-ik_2} + g_{-1}(t, k_c)e^{ik_2}], 
\end{equation}  
with which we can extract an effective pseudo-magnetic field $\vec B_{\rm eff}$, such that $G_{\rm eff}(t, k_c, k_2) = \vec B_{\rm eff}(t, k_c, k_2) \cdot \vec \sigma$. 
If we recast
\begin{equation}
g_{0} = \vec{n}_{0} \cdot \vec{\sigma}, \quad g_{\pm 1} = \frac{\vec{n}_{1} \pm i  \vec{n}_{2}}{2} \cdot \vec{\sigma},
\label{eq:n012}
\end{equation} 
we immediately find 
\begin{equation}
\vec B_{\rm eff} = \vec{n}_{0} + \vec{n}_{1} \cos(k_2) + \vec{n}_{2} \sin(k_2).
\end{equation}

Therefore, to solve for $k_c$, we only need to study $\vec B_{\rm eff}(t, k_c, k_2)$, whose tip traces an ellipse as $k_2$ varies. 
Compared to the complex trace of $\hat B(t, k_c, k_2)$ in Eq.~(\ref{eq_Bt}), 
we have effectively filtered out the fast-varying components. 
We can follow the reasoning in the equilibrium case and assert that at $k_c$ we have similarly
\begin{equation}
\langle \beta_c \vert \vec B_{\rm eff}(t, k_c, k_2) \cdot \vec \sigma \vert \beta_c \rangle = 0.
\end{equation}
We can identify the coplanarity condition of the pseudospin magnetic field $\vec{B}_{\rm eff}$ as
the coplanarity of $\vec{n}_{0}$, $\vec{n}_{1}$, and $\vec{n}_{2}$, i.e., 
\begin{equation}
\label{eq:n012t}
\vec{n}_0 \cdot (\vec{n}_1 \times \vec{n}_2) = 0. 
\end{equation}
After straightforward algebras, we arrive at the following triple-integral condition
\begin{align}
    \label{eq_linear_diag}
        & \int_0^{2\pi}dk_2 \int_0^{2\pi}dk_2^{\prime} \int_0^{2\pi}dk_2^{\prime\prime} \, h(k_2, k_2^{\prime}, k_2^{\prime\prime}) \nonumber \\
        & \times \left \{ \left[\hat{B}(t, k_c,k_2)\times\hat{B}(t, k_c,k_2^{\prime})\right]\cdot\hat{B}(t, k_c,k_2^{\prime\prime}) \right \} =0,
\end{align}
where $h(k_2, k_2^{\prime}, k_2^{\prime\prime}) = \sin(k_2-k_2^{\prime})+\sin(k_2^{\prime}-k_2^{\prime\prime})+\sin(k_2^{\prime\prime}-k_2)$
and $\hat{B}(t, k_1, k_2)$ is defined in Eq.~(\ref{eq_Bt}) for the dynamical case. 
The result can be regarded as the relaxation of the equilibrium coplanarity condition at $k_c$ 
to the non-equilibrium case. 
The stronger coplanarity condition in Eq.~(\ref{eq:coplanarity}) reincarnates now 
as a time-dependent factor in the integrand, which is modulated by a static geometrical factor
that depends only on momenta, but not on the band structure.

%%%%%%%%%%%%%%%%%%%%%%%%%%%%%%%%%%%%%%%%%%%%%%%%%%%%%%%%%%%%%%%%%%%%%%%%%%%%%%%%%%%%%%%%%%%%%%%%%
\subsection{Dynamics of the entanglement crossing momentum}
\label{sec:crossing_dynamics}

\begin{figure}
    \includegraphics[width=8cm]{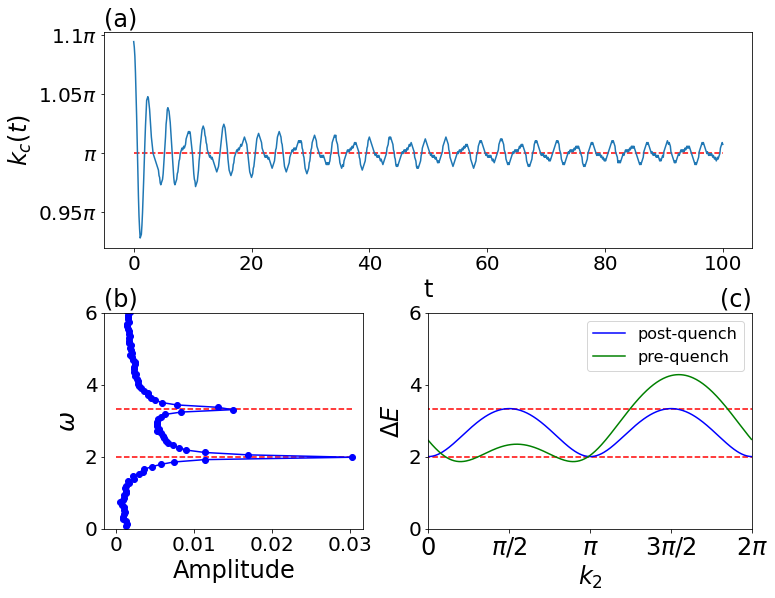}
    \caption{\label{fig:ES_kc}
    (a) Evolution of $k_c(t)$ in the system with $L_2 = 400$ 
    after the quench from $M = \sqrt{3}/3$ to $M = 0$ with fixed $\phi = \pi/2$. 
    $k_c(t)$ oscillates around the post-quench equilibrium $k_c^{\prime}$ (blue dashed line). 
    (b) The corresponding frequency spectrum of $k_c(t)$. 
    (c) The interband excitation energy $\Delta E$ at the corresponding crossing points 
    for the pre- (green curve) and post-quench (blue curve) Hamiltonians. 
    The dashed lines mark the post-quench $\Delta E = 2 \vert \vec{B}' \vert$ 
    of the van Hove singularities in (c), 
    which coincide with the spectral peaks in (b). }
\end{figure}

Let us come back to the example in Fig.~\ref{fig:geometry} and explore in detail the dynamics after the system quenches from $M = \sqrt{3}/3$ to $M = 0$ at $t = 0$ with fixed $\phi = \pi/2$. 
We determine $k_c(t)$ by Eq.~(\ref{eq_linear_diag}) and plot its evolution in Fig.~\ref{fig:ES_kc}(a) 
for $0 < t < 100$. 
The entanglement spectrum crossing shows decaying oscillations 
around the post-quench $k_c^{\rm post} = \pi$,
starting from the pre-quench $k_c^{\rm pre} = \pi + \sin^{-1} (\sqrt{3}/6)$ at $t = 0$. 
The Fourier transformation of $k_c(t)$ exhibits two sharp peaks in the frequency space,
as shown in Fig.~\ref{fig:ES_kc}(b).
The values of the peaks coincide with the extrema of the post-quench interband excitation energy 
$\Delta E = 2 \vert \vec{B'(k_1, k_2)} \vert$ at $k_1 = k_c^{\rm post}$, as illustrated in 
Fig.~\ref{fig:ES_kc}(c).
For comparison, we also plot the pre-quench interband excitation energy, whose extrema 
match three different frequencies. 
The observation implies that the interband excitations play an important role in the quench dynamics, 
especially near the extrema where the density of states is singular. 

To understand the dependence of $k_c$ on $t$, we need to evaluate $g_{0}(t, k_1)$ and $g_{1}(t, k_1)$, which are defined in Eq.~(\ref{eq:Gntk1}).
The results from solving Eq.~(\ref{equation:C_gamma=2}) agree very well with 
those directly obtained from the numerical calculation of the OPES without approximations, 
which implies that the edge-state ansatz wave function and 
the tridiagonal approximation of the correlation matrix are well justified. 
As $k_2$ varies, the post-quench pseudospin magnetic field $\hat B'$ oscillates, whose magnitude enters the frequency 
of the dynamical pseudospin magnetic field $\hat{B}(t, k_1, k_2)$. 
For sufficiently large $t$, the phase of this oscillating factor varies rapidly
except in the neighborhood of the stationary points of $\hat B'$, 
at which $\partial \vert \vec{B}' \vert/\partial k_2 = 0$.
There are four stationary points for a given $k_1$ in the Haldane model [Eq.~(\ref{eq_B})], two with positive 
curvatures, which we label as $k_{1,+1}$ and $k_{2,+1}$, the other two with negative ones, which we label as $k_{1,-1}$ and $k_{2,-1}$.
They correspond to the locations of the van Hove singularities in the density of states, 
and the corresponding frequency $2 \vert \vec{B}' \vert$ is the interband excitation energy for the post-quench Hamiltonian. 
In the following, we use the stationary phase approximation to analyze the dominant contributions near $k_{2s} = k_{i,\sigma}$ with $i = 1,2$ and $\sigma = \pm 1$. 

Starting from Eq.~(\ref{eq:Gntk1}), we can first separate the correlation matrix into a steady-state contribution 
and a time-dependent one:
\begin{eqnarray}
g_{n}(t, k_1) &=& \frac{1}{2\pi} \int_0^{2\pi} e^{ink_2}  \left [ \hat{B}(t, k_1, k_2) \cdot \vec{\sigma} \right ] dk_2 \nonumber \\
&=& g^{\infty}_{n}(k_1) + \delta g_{n}(t, k_1),
\end{eqnarray}
where we have
\begin{equation}
g^{\infty}_{n}(k_1) = \frac{1}{2\pi} \int_0^{2\pi} e^{ink_2}  \left (\vec{b}_3 \cdot \vec{\sigma} \right ) dk_2
\end{equation}
and 
\begin{eqnarray}
\delta g_{n}(t, k_1) = \frac{1}{2\pi} \int_0^{2\pi} && e^{ink_2}  \left [ \cos(2\vert \vec{B}' \vert t) (\vec{b_1} \cdot \vec{\sigma}) \right . \nonumber \\
+ &&\left . \sin(2\vert \vec{B}' \vert t) (\vec{b_2} \cdot \vec{\sigma}) \right ] dk_2.
\label{eq:deltaG}
\end{eqnarray}
Alternatively, we expect that, according to definitions in Eq.~(\ref{eq:n012}), $\vec{n}_i$ can also be separated into steady-state and time-dependent contributions 
\begin{equation}
\vec{n}_i(t,k_1) = \vec{n}_i^{\infty}(k_1) + \delta \vec{n}_i(t,k_1), \quad i = 0,1,2.
\end{equation}
In particular, we have $\hat B (t = 0) = \vec{b}_1 + \vec{b}_3 = \hat B$ at $t = 0$, so $k_c (t = 0) = k_c^{\rm pre}$ is 
simply the momentum of the entanglement zero modes of the pre-quench Hamiltonian. 
In the long time limit ($t \gg 1$), however, the time-dependent terms in $\hat{B}(t)$ oscillate so rapidly that 
$\delta \vec{n}_i$s vanish. 
Effectively, we have $\hat B (t \gg 1) = \vec{b}_3$, whose unit vector for each $(k_1, k_2)$ pair is identical to $\hat{B}'(k_1, k_2)$, 
so  $k_c$ approaches $k_c^{\rm post}$, the momentum of the entanglement zero modes 
of the post-quench Hamiltonian.
In other words, we expect 
\begin{equation}
\vec{n}_0^{\infty}(k_c^{\rm post}) \cdot [\vec{n}_1^{\infty}(k_c^{\rm post})  \times \vec{n}_2^{\infty}(k_c^{\rm post})]  = 0.
\end{equation}
In practice, the time-dependent terms are small compared with the steady-state contributions for not too small $t$, so the time-dependent zero-mode momentum $k_c(t)$ varies in the vicinity of $k_c^{\rm post}$, 
where we can linearize the triple product,
\begin{equation}
\vec{n}_0^{\infty} (k_1) \cdot [\vec{n}_1^{\infty} (k_1)  \times \vec{n}_2^{\infty} (k_1) ]  \propto \left (k_1 - k_c^{\rm post} \right ).
\end{equation}
In general, $k_c(t)$ can be calculated from Eq.~(\ref{eq:n012t}) in which we can expand $\vec{n}_i(t,k_1)$ and neglect the high-order terms in $\delta \vec{n}_i$ and arrive at 
\begin{eqnarray}
k_c(t) - k_c^{\rm post} &\propto& \vec{n}_0^{\infty} \cdot (\vec{n}_1^{\infty}  \times \vec{n}_2^{\infty}) \nonumber \\
&\approx& - \delta \vec{n}_0 \cdot (\vec{n}_1^{\infty} \times \vec{n}^{\infty}_2)   - \delta \vec{n}_1 \cdot (\vec{n}_2^{\infty} \times \vec{n}^{\infty}_0)   \nonumber \\
&& - \delta \vec{n}_2 \cdot (\vec{n}_0^{\infty} \times \vec{n}^{\infty}_1),  
\end{eqnarray}
where all $\vec{n}_i^{\infty}$ and $\delta \vec{n}_i$ are evaluated at $k_1 = k_c(t)$.

\begin{figure}
    \includegraphics[width=8cm]{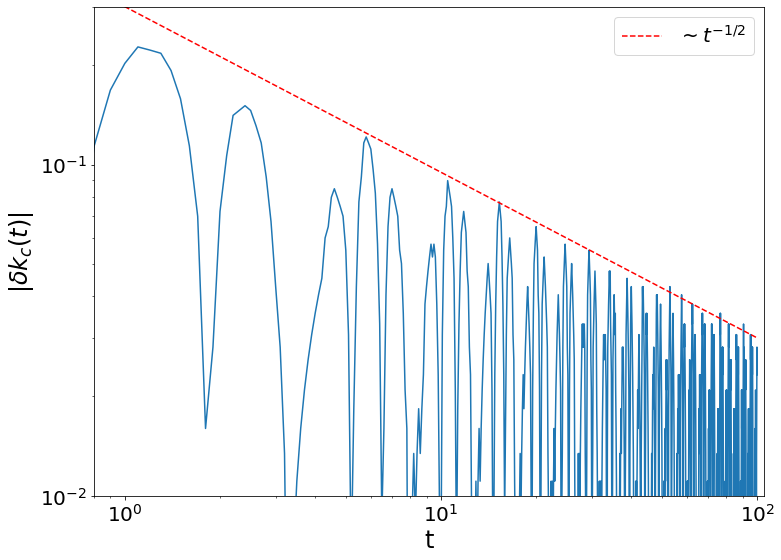}
    \caption{\label{fig:kctrend}
    Evolution of $k_c(t)$ in Fig.~\ref{fig:ES_kc}(a) replotted as $\vert \delta k_c(t) \vert = \vert k_c(t) - k_c^{\rm post} \vert$ on a double-logarithmic scale. The red dashed line, which goes as $t^{-1/2}$, is a guide to the eye demonstrating the decaying trend of the $k_c(t)$ deviation. 
    }
\end{figure}

We can proceed by expanding $\hat B'(k_1, k_2)$ around the saddle points $k_2 = k_{i,\sigma}$ for given $k_1$, such that
\begin{equation}
e^{2 i \vert \vec{B}' \vert t} \approx e^{2 i \vert \vec{B}' (k_1, k_{i,\sigma}) \vert t}  \exp \left [ i \mu_{i,\sigma}  (k_2 - k_{i,\sigma})^2 t \right],
\end{equation}
where
$\mu_{i,\sigma} = \partial^2 \vert \vec{B}' \vert/\partial k_2^2 \vert_{k_{2} = k_{i,\sigma}}$.
Both the real and imaginary parts of $\delta \vec{n}_i (t, k_1)$ can then be calculated by performing Gaussian integrals by noticing 
\begin{equation}
\cos(2\vert \vec{B}' \vert t) \vec{b}_1 + \sin(2\vert \vec{B}' \vert t) \vec{b}_2 = 
\Re \left [e^{2 i \vert \vec{B}' \vert t} \vec{b}_{-} \right ],
\end{equation}
where $\vec{b}_{-}  = \vec{b}_1 - i \vec{b}_2$. 
In the stationary phase approximation, 
the leading contributions to the integral of $\delta \vec{n}_i$ thus come from 
the stationary points $k_2 = k_{i,\sigma}$ with $i = 1,2$ and $\sigma = \pm 1$.
They oscillate with frequency being the corresponding energy gap 
$2 \vert \vec{B}' (k_1, k_{i,\sigma}) \vert$ at different $k_{2s}$
with a common decay of $t^{-1/2}$, resulting from the Gaussian integrals. 
Accidentally, the corresponding integrands in $\delta \vec{n}_i$ may vanish as $(k_2 - k_{2s})^q$ 
near the saddle points, in which case $\delta \vec{n}_i \sim t^{-(q+1)/2}$. 
This, however, does not happen for all $\delta \vec{n}_i$, so the leading correction in $k_c(t) - k_c^{\rm post}$ is still dominated by $t^{-1/2}$.
Therefore, the zero-mode momentum $k_c(t)$ starts from $k_c^{\rm pre}$ at $t = 0$ and oscillates around $k_c = k_c^{\rm post}$ with the amplitude decreasing as $t^{-1/2}$, and
the dominant frequencies of the oscillations correspond to the interband excitation energies 
at the stationary points. 
Figure~\ref{fig:kctrend} plots $\vert \delta k_c(t) \vert = \vert k_c(t) - k_c^{\rm post} \vert$ 
on a double-logarithmic scale. 
The peaks of the curve are roughly bound by the red dashed line, which goes as $t^{-1/2}$ and is a guide to the eye. 
This demonstrates that the oscillations are dominated by the excitations around stationary points. 

%%%%%%%%%%%%%%%%%%%%%%%%%%%%%%%%%%%%%%%%%%%%%%%%%%%%%%%%%%%%%%%%%%%%%%%%%%%%%%%%%%%%%%%%%%%%%%%%%
\section{Dynamics of the entanglement velocity}
\label{sec:geometry}

So far, we have discussed the quantum quench within the same topological phase, 
and the pre- and post-quench equilibrium states are characterized by 
different entanglement crossing momentum $k_c$. 
As a result, the dynamical crossing $k_c(t)$ starts from 
the pre-quench $k_{c}^{\rm pre}$ and oscillates around the post-quench $k_{c}^{\rm post}$. 
This seems to suggest that the dynamics not associated with the decaying 
oscillatory behavior of $k_c$ is trivial, e.g., when 
the pre- and post-quench equilibrium states share the same $k_c$.
This happens, in particular, when we fix $M = 0$ before and after the quench, 
such that $k_c^{\rm post} = k_c^{\rm pre} = \pi$. 

This cannot be correct because if we increase the distance 
between the post- and pre-quench states in the phase diagram,  
we eventually arrive at a sudden quench from one topological phase (say, $C = 1$) 
to another ($C = -1$) with fixed $M = 0$.
Even though $k_c$ is not changed in this case, the sign change of the topological index indicates
that the equilibrium edge modes on the same open boundaries change their directions accordingly. 
It is then an intriguing question to ask what dynamical behavior is the precursor of 
edge-mode direction change in the same-phase ($C = 1$ to 1) quenches along $M = 0$. 
Such a connection to the corresponding topological quench is natural 
as the dynamics of $k_c(t)$ discussed so far can be thought of as the precursor of 
the topological transition from the $C = 1$ phase to the $C = 0$ phase. 
In the latter, the solution of Eq.~(\ref{eq:coplanarity2}) yields a $k_c$ 
that violates Eq.~(\ref{eq:topology}) for nontrivial topology to exist. 
This points us to the change of slope or velocity of the edge mode 
in the entanglement spectrum. 

As we mentioned in the equilibrium case at the end of Sec.~\ref{sec:eq_opes}, the entanglement crossing momentum $k_c$ is not complete in describing the edge modes in the vicinity of the crossing, hence distinguishing the topological states. 
It can be complemented, e.g.,  by the entanglement velocity, i.e., the slope of the edge modes at the crossing,
\begin{equation}
v = \left . \frac{\partial \epsilon}{\partial k_1} \right \vert_{\epsilon =0} = 4 \left . \frac{\partial \lambda}{\partial k_1} \right \vert_{\lambda=\frac12}.
\label{eq:velocity}
\end{equation}
For simplicity, we consider quench far from the phase boundary such that the edge modes decay sufficiently fast.  
In the tridiagonal approximation in Eq.~(\ref{equation:C_gamma=2}), we can make a first-order perturbation in the vicinity of $k_c$ to 
calculate the edge-mode entanglement energy by 
\begin{equation}
\lambda_{\rm edge} (t, k_1) = \frac12 \pm \frac12 \langle \beta_c \vert g_0(t,k_1) \vert \beta_c \rangle,
\end{equation}
where
\begin{eqnarray}
g_{0}(t, k_1) = \frac{1}{2\pi} \int_0^{2\pi} && \left [ \vec{b}_3 \cdot \vec{\sigma} + \cos(2\vert \vec{B}' \vert t) (\vec{b_1} \cdot \vec{\sigma}) \right . \nonumber \\
+ &&\left . \sin(2\vert \vec{B}' \vert t) (\vec{b_2} \cdot \vec{\sigma}) \right ] dk_2,
\label{eq:deltaG0}
\end{eqnarray}
with $\vec{b}_i$s defined in Eqs.~(\ref{eq:b1})-(\ref{eq:b3}). 
The dependence of $g_0(t, k_1)$ on $k_1$ 
is encoded explicitly in the effective pseudospin magnetic field.
As discussed in Sec.~\ref{sec:crossing_dynamics},
$g_0$ can be split into two terms, corresponding to the steady-state 
and time-dependent contributions. 
Accordingly, we can write 
\begin{equation}
v(t) = v^{\infty} + \delta v(t), 
\end{equation}
where 
\begin{eqnarray}
v^{\infty} &=& 2 \left \langle \beta_c \left \vert \frac{\partial \,g^{\infty}_0}{\partial k_1}  \right \vert \beta_c \right \rangle, \\
\delta v(t) &=& 2 \left \langle \beta_c \left \vert \frac{\partial \delta g_0}{\partial k_1}  \right \vert \beta_c \right \rangle.
\end{eqnarray}
For sufficiently large $t$, we notice the leading contribution in $\delta v(t)$ comes from the derivatives 
of the fast oscillating $\cos(2\vert \vec{B}' \vert t)$ and $\sin(2\vert \vec{B}' \vert t)$ 
with respect to $k_1$, which yield an explicit factor of $t$. 
Combined with the additional factor of $t^{-1/2}$ from the stationary-phase approximation, 
we expect that the envelope of the oscillation of $\delta v(t)$ goes as $t^{1/2}$ as $t$ increases. 
The prefactor increases with the difference between the post- and pre-quench states. 

\begin{figure}
    \includegraphics[width=8cm]{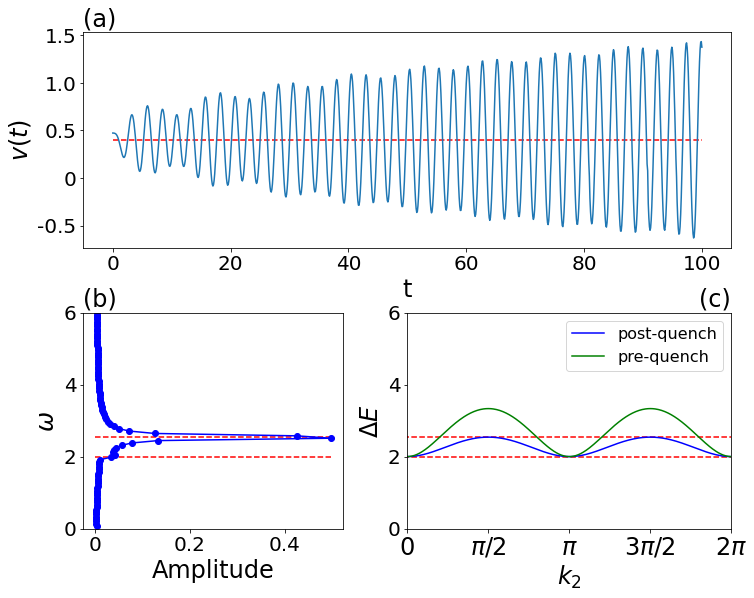}
    \caption{\label{fig:vkc}
    (a) Evolution of the entanglement velocity $v(t)$, which is obtained from the entanglement spectrum in the absence of the tridiagonal approximation, according to Eq.~(\ref{eq:velocity}). The system has $L_2 = 400$, and the quench is from $\phi = \pi/2$ to $\phi = \pi/5$ with fixed $M = 0$. 
    (b) The Fourier transform of $v(t)$.
    (c) The interband excitation energy $\Delta E$ at the crossing momentum $k_c = \pi$
    for the pre- (green curve) and post-quench (blue curve) Hamiltonians. 
    The spectral peaks in (b) agree with the extremal interband excitation energies of the post-quench Hamiltonian. }
\end{figure}

For a concrete example, we consider an $L_2 = 400$ system quenching from 
$\phi = \pi/2$ to $\phi = \pi/5$ with fixed $M = 0$, in which case
a pair of zero modes persists at $k_c = \pi$ where we can numerically calculate 
the entanglement velocity $v(t)$ from the entanglement spectrum 
in the absence of the tridiagonal approximation. 
As shown in Fig.~\ref{fig:vkc}(a), the velocity $v(t)$ develops rapid oscillations 
whose amplitude increases with $t$. 
The red dashed line indicates the corresponding velocity $v_{\rm inf}$ if we set $\vec{B}(t) = \vec b_3$ in Eq.~(\ref{eq_Bt}), which can be regarded as the long-time limit after the oscillations die out in the presence of dissipations. 
As we have analyzed for $k_c(t)$, we calculate the frequency spectrum of $v(t)$ and find that it is dominated by the modes near the maximum of the excitation gap for the post-quench Hamiltonian,
as shown in Fig.~\ref{fig:vkc}(b) and (c). 
At the minimum of the gap, however, we only find a shoulder in the spectrum. 
Following the same arguments for $k_c(t)$ in Sec.~\ref{sec:crossing_dynamics}, 
we can understand that the velocity oscillations reflect, again, the interband excitations, 
because the prefactor of the time dependence is dominated by 
the energy gap $2 \vert \vec{B}' (k_1, k_{i,\sigma}) \vert$ at the four stationary points. 

The sign of the velocity is important. 
The bipartition in this study creates two artificial edges, along each of which lives a branch of edge modes. 
Together, they manifest a zero-mode crossing in the entanglement spectrum and
correspond to two velocities of opposite signs. 
To study the evolution of the velocity, we must fix our focus on a single edge and 
monitor the change of the slope near the corresponding zero mode. 

\begin{figure}
    \includegraphics[width=8cm]{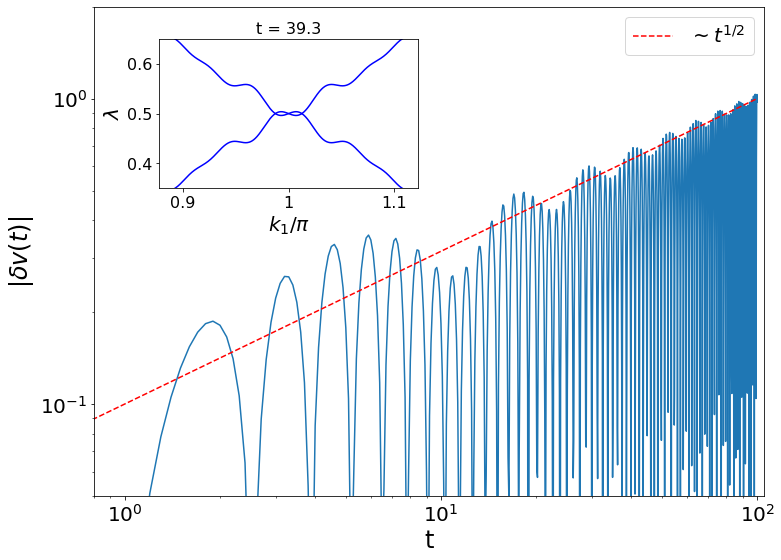}
    \caption{\label{fig:vtrend}
    Evolution of $v(t)$ in Fig.~\ref{fig:vkc}(a) replotted as $\vert \delta v(t) \vert = \vert v(t) - v_{\rm \inf} \vert$ on a double-logarithmic scale, 
    where $v_{\rm inf}$ can be regarded as the long-time average or limit. The red dashed line, which goes as $t^{1/2}$, 
    is a guide to the eye for the amplitude increasing trend. 
    The inset shows the dynamical OPES at $t = 39.3$ near $k_1 = \pi$ for the two edge modes. 
    Three consecutive crossings are visible, and, accordingly, the entanglement velocity changes sign near the center.  
    }
\end{figure}

Fig.~\ref{fig:vtrend} plots the velocity oscillations $\vert \delta v(t) \vert = \vert v(t) - v_{\rm \inf} \vert$ on a double-logarithmic scale. 
As we discussed above, the curve is roughly bound by a trend that goes as $t^{1/2}$.
In addition, the oscillations show a beating pattern, which is consistent with the energy difference of 
the two extremal excitation gaps. 
The increasing amplitude in the velocity oscillations guarantees its sign change after sufficiently long time in a large enough sample,
which indicates a reconstruction of the edge mode and the appearance of counterpropagating edge states. 
As a result, additional entanglement crossings appear, as shown in the OPES at $t = 39.3$, 
at which the sign change can be observed in Fig.~\ref{fig:vkc}(a). 
The additional crossings appear in pairs around $k_c = \pi$, 
hence preserve the nontrivial topology of the system. 

\begin{figure}
    \includegraphics[width=9cm]{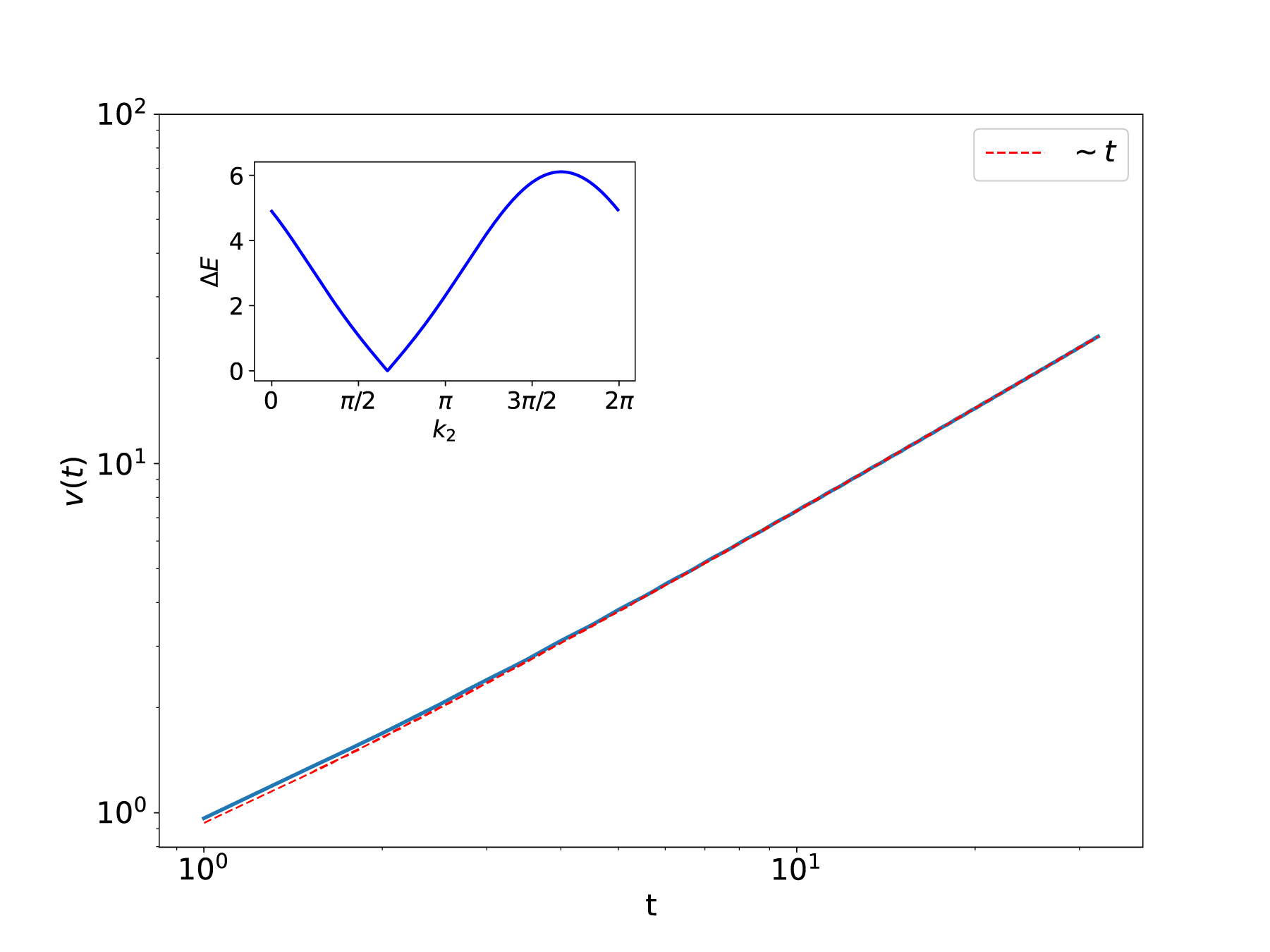}
    \caption{\label{fig:crit}
Evolution of $v(t)$ for quenching to the critical case with the Hamiltonian parameters $M=\sqrt{3}-0.1$ to $M=\sqrt{3}$ with fixed $\phi=\pi/2$, on a double-logarithmic scale. Red dashed line shows the linear behavior. Inset: Interband excitation spectrum of the post-quench Hamiltonian with the crossing momentum $k_1=k_c^{\rm post}$.}
\end{figure}

Our analysis can be readily extended to quenches to the critical case, with the post-quench Hamiltonian 
located at the boundary between phases of different Chern index. 
As an example, we quench $M=\sqrt{3}-0.1$ to $M=\sqrt{3}$ with fixed $\phi=\pi/2$. The interband excitation energy exhibits a Dirac cone at $(k_1,k_2)=(k_c^{\rm post}, 2\pi/3)$ with the crossing momentum $k_c^{\rm post}=4\pi/3$ (inset of Fig.~\ref{fig:crit}).
%The velocity at $k_1 = k_c$ vanishes except for the Dirac point, so 
The saddle-point contribution of $t^{-1/2}$ disappears, and the long-time behavior of the entanglement velocity is dominated by the Dirac cone and increases as $\sim t$, as shown in Fig. 10. 
There are no rapid oscillations in $t$, because the dynamics at the Dirac point is simply determined by $\Delta E = 0$. 
%On the other hand, the $\sim t^{1/2}$ time dependence from the remaining saddle point 
%appears as short-time corrections. 
Analysis for the dynamics of the momentum crossing shows that $k_c$ saturates at $4\pi/3$ in the long-time limit, 
with a $\sim t^{-1}$ correction from the Dirac cone. % and a $\sim t^{-1/2}$ correction from the saddle point.
% yields a $\sim t^{-1}$ dependence from the Dirac cone, and $\sim t^{-1}$ dependence from the Dirac cone,
%which means $k_c(t)$ is saturating at $4\pi/3$ in the long time limit.
%Therefore, for quenches to the critical points, the long-time behavior of the crossing momentum and the entanglement velocity is dominated by the gapless Dirac structure in the spectrum.
We can regard this results as an extreme case of Figs.~\ref{fig:ES_kc} and \ref{fig:vkc}(a), 
in which the Fourier peaks are shifted to $\omega = 0$, such that the oscillations disappear as their period becomes infinitely long and the asymptotic behavior of the envelop functions takes over. 
Correspondingly, the change in the saddle point structure at criticality leading to different exponent has been studied for other observables, see e.g., Refs. \cite{Makki22, Sim22, Ramos23, Cao24}.

%%%%%%%%%%%%%%%%%%%%%%%%%%%%%%%%%%%%%%%%%%%%%%%%%%%%%%%%%%%%%%%%%%%%%%%%%%%%%%%%%%%%%%%%%%%%%%%%%
\section{Evolution of Dynamical Winding Vector}
\label{sec:winding}

In Fig.~\ref{fig:ES2} we showed that the entanglement velocity $v(k_c)$ complements the zero-mode momentum $k_c$ in describing different topological states with the same Chern index. 
While it is tempting to use the combination of $k_c$ and $v(k_c)$ to describe the dynamical evolution 
of the topological state, the distinct time dependence of the two variables hints that they cannot 
be put on equal footing in the non-equilibrium case. 
In particular, the amplitude of the oscillations in $v$ increases without bound in the long-time limit. 
Such oscillations are associated with particles exciting between the two bands and 
are dominated by the stationary points of the excitation gap. 
Thus, the geometrical information encoded in the velocity is expected to be washed out gradually
by the ever increasing oscillations. 
Besides, we can obtain $v(k_c)$ only after we calculate $k_c$ at  a give time; 
this also makes the two variables not compatible. 
We, therefore, turn our attention to the winding vector $\hat m(k_c)$ in the equilibrium case and 
study its counterpart in the non-equilibrium case. 

In the equilibrium case, $\hat m(k_c)$ represents the plane on which the pseudospin magnetic field 
vector $\vec B(k_c, k_2)$ winds. 
In particular, $\hat m(k_c)$ is perpendicular to both $\vec n_1$ and $\vec n_2$ that span the plane. 
In the non-equilibrium case, however, $\vec B(t, k_c, k_2)$ does not wind on a plane. 
But, as we have shown in Sec.~\ref{sec:noneq_cond}, after filtering away high-frequency components, we recover the equilibrium 
decomposition of $\vec B_{\rm eff}(t, k_c, k_2)$ into time-dependent $\vec n_0$, $\vec n_1$, and $\vec n_2$, 
which allow us to define the non-equilibrium $\hat m(k_c)$, again, as 
\begin{equation}
\hat m(k_c) = \frac{1}{2\pi} \int \hat B_{\rm eff} \times d\hat B_{\rm eff} 
= \frac{\vec{n}_1 \times \vec{n}_2}{\vert \vec{n}_1 \times \vec{n}_2 \vert}.
\end{equation}

\begin{figure}[t]
    \includegraphics[width=8cm, trim=0 0.8cm 0 0]{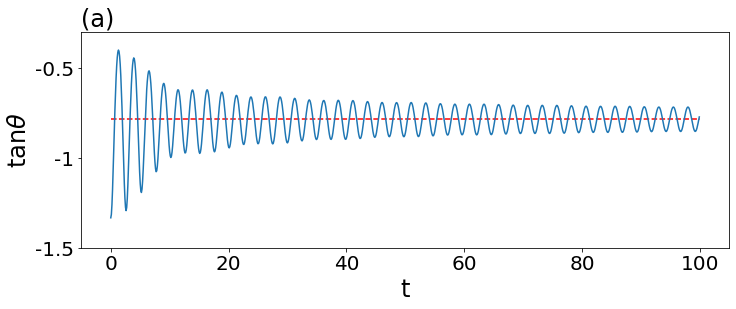}
    \includegraphics[width=3.5cm, trim=0 -1cm 0 0]{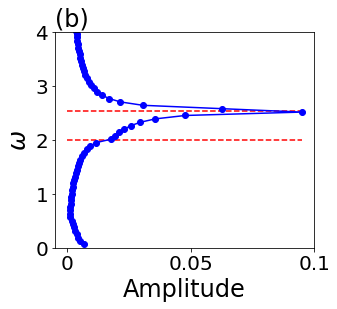}
    \includegraphics[width=4cm]{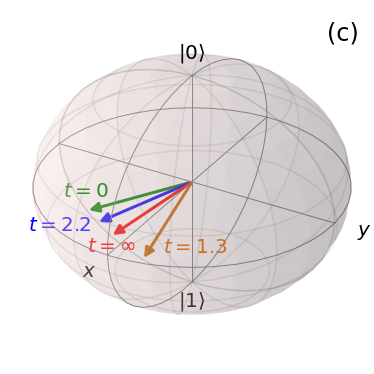}
        \caption{\label{fig:tantheta}
    (a) Evolution of $\tan \theta$ for the tilting angle $\theta(t)$ from basis vector $\ket{0}$ 
    of the winding vector as a function of $t$. 
    The system has $L_2 = 400$, and the quench is from $\phi = \pi/2$ to $\phi = \pi/5$ with fixed $M = 0$.
    The red dashed line indicates the value of $\tan \theta$ in the long-time limit. 
    (b) The Fourier transform of $\tan \theta(t)$. The two red dashed lines indicate the minimum and maximum excitation gaps. 
    (c) Representations of the winding vectors at $t = 0$, 1.3, 2.2, and $t = \infty$ on a Bloch sphere. The winding vector swings back-and-forth across its long-time limit.  
    }
\end{figure}

Let us return to the example in Sec.~\ref{sec:geometry}, in which the system quenches from 
$\phi = \pi/2$ to $\phi = \pi/5$ with fixed $M = 0$. 
In this case, a pair of zero modes persists at $k_c = \pi$ 
and $\hat m(k_c)$ is confined within the $y$-$z$ plane. 
We define $\tan \theta = m_y (k_c) / m_z (k_c)$ and plot it as a function of time 
in Fig.~\ref{fig:tantheta}(a). 
We find that $\tan \theta$ oscillates with a decreasing amplitude, whose spectral analysis in Figure~\ref{fig:tantheta}(b)
shows a similar peak and shoulder as in Fig.~\ref{fig:vkc}(b) for the entanglement velocity. 
The amplitude decays in a power law with exponent $-1/2$, the same as that of $k_c$. 
In Fig.~\ref{fig:tantheta}(c) we illustrate the oscillation of $\hat m(k_c)$ in the $y$-$z$ plane 
by plot it with the Bloch sphere at several time, including $t = 0$ [the initial $\hat m(k_c)$ with 
$\phi = \pi/2$] and $t = \infty$ [the final $\hat m(k_c)$ with $\phi = \pi/5$]. 

Generally, a complete description of the dynamical evolution of 
the system requires the specification of all $\vec B(k_1, k_2)$. 
However, in this example we have shown that it is adequate to describe 
the non-equilibrium topological system by specifying
a winding vector $\hat m(k_c)$ that defines a great circle on the Bloch sphere, 
around which $\vec B_{\rm eff}(t, k_1 = k_c(t), k_2)$ winds, 
where $k_c$ is the generic zero-mode momentum 
associated with the nontrivial topology of the dynamical system. 
In this case, $\hat m(k_c)$ starts from the equilibrium value of the pre-quench Hamiltonian
and approaches that of the post-quench Hamiltonian in an oscillatory fashion. 
The amplitude of the oscillations follows a power law with exponent $-1/2$, indicating that
the range of constructive interference shrinks with time linearly.
In this example we fix $M = 0$ for both pre- and post-quench Hamiltonians 
so the chiral symmetry persists in the evolution. 
As a result, a single parameter, e.g. $\tan \theta$, is sufficient to describe the dynamics. 
The spectral analysis of the parameter allows us to find out the interband excitation gap edges 
and their corresponding weights. 

\begin{figure}
    \includegraphics[width=3.5cm]{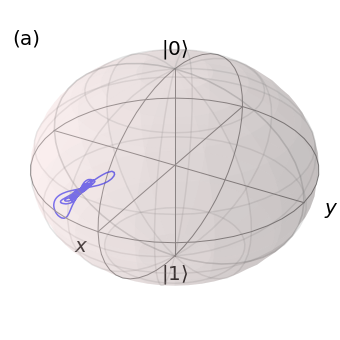}
    \includegraphics[width=4cm]{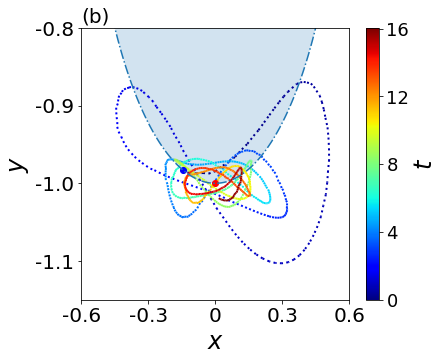}
        \caption{\label{fig:mtrace}
    (a) Evolution of the arrow head of the winding vector $\hat m(t)$ on the Bloch sphere for $0 < t < 16$. 
    The system has $L_2 = 400$, and the quench is from $M = \sqrt{3}/3$ to $ M = 0$ with $\phi = \pi/2$.
    (b)  The stereographic mapping of the head of $\hat m(t)$ on the Bloch sphere. Color, from blue to red, labels time after the quench. The blue and red dots corresponds to the winding vectors of the pre- and post-quench Hamiltonians, respectively. The blue shadowed region surrounded by the dotted-dash 
line corresponds to the end point of $\hat m(k_c)$ for the $C = 1$ topological phase, as illustrated in Fig.~\ref{fig:projection}.
    }
\end{figure}

More complex examples arise as the chiral symmetry is broken in either pre- or 
post-quench Hamiltonian, or both.
For example, let us revisit the example in Sec.~\ref{sec:crossing_dynamics}, 
where the quench is from $M = \sqrt{3}/3$ to $ M = 0$ with $\phi = \pi/2$.
In this case, we need to first calculate $k_c$ and then define $\hat m(k_c)$ through the 
Fourier transform of $\vec B_{\rm eff}(t, k_1 = k_c(t), k_2)$. 
Fig.~\ref{fig:mtrace}(a) illustrates the evolution of the head of $\hat m(k_c)$ 
on the Bloch sphere. 
The arrow head appears to trace smaller and smaller figure-eight--like knots, indicating that 
$\hat m(k_c)$ converges again to that of the post-quench Hamiltonian in the long-time limit. 
The complex trace can again be illustrated by the stereographic projection to the plane 
tangent at the south pole (i.e., state $\ket{1}$) of the Bloch sphere, 
as shown in Fig.~\ref{fig:mtrace}(b) where time is indicated by the color map.
For clarity, we only showed the portion for $t < 16$. 
The projected equilibrium $\hat m(k_c)$ for the pre- and post-quench Hamiltonians are 
represented by the blue and red dots, respectively. 
They are located on boundary of the blue-shadowed region, which corresponds to the 
set of equilibrium topological states with $C = 1$, as illustrated in Fig.~\ref{fig:projection},
even though they are not on the phase boundary in the phase diagram of $(M, \phi)$ 
in Fig.~\ref{fig:model}.

Surprisingly, the size of the trace is much larger than the distance 
between the pre- and post-quench points, even if we take into account the distortion 
by the stereographic projection. 
Furthermore, a significant portion of the trace features a winding vector that points to the direction 
not matched by its equilibrium counterpart, unlike in the previous example illustrated in Fig.~\ref{fig:tantheta}. 
In other words, the restriction on the winding of the pseudospin magnetic field related to the equilibrium 
band structure is relaxed in the non-equilibrium case due to the nontrivial time-dependent geometry 
of the effective pseudospin magnetic field. 
This is quite unexpected especially when the pre- and post-quench Hamiltonians are located deep in the topological phase. 

%%%%%%%%%%%%%%%%%%%%%%%%%%%%%%%%%%%%%%%%%%%%%%%%%%%%%%%%%%%%%%%%%%%%%%%%%%%%%%%%%%%%%%%%%%%%%%%%%%%%
\section{Summary and Discussion}
\label{sec:summary}

The non-equilibrium quench dynamics of isolated quantum systems has generated tremendous 
amount of theoretical interest on phenomena such as the linear increase of entanglement entropy, light-cone spreading of quasiparticles and their correlations, as well as phenomena of persistent oscillations and fluctuations due to many-body effects \cite{Lazarides14, Kiendl17, Kormos17, CAlvaredo20, Serbyn21, Delfino22, Robertson24}. 
In topological systems additional interest has been given to the edge modes that correspond to 
the nontrivial topology of the bulk. 
In this study we focus on describing non-equilibrium topological states by additional 
geometrical degrees of freedom that are associated with the edge modes in the entanglement spectrum, 
with examples in Haldane's honeycomb model. 

In the equilibrium case we confirm the criterion for nontrivial topology 
in the entanglement spectrum is the existence of zero modes propagating along the edges, 
or a crossing of the edge modes at half entanglement occupation.
At the characteristic momentum $k_c$, the normalized pseudospin magnetic field $\hat B(k_c, k_2)$ 
sweeps a great circle on the Bloch sphere. 
The normal direction of the circle $\hat m(k_c)$ can be related to the wave function of the zero modes. 
The unit vector, fixed by two parameters, encodes the geometrical information 
beyond the topological Chern index of the nontrivial state. 
So the family of topological states sharing the same Chern index can be distinguished 
by their $\hat m(k_c)$ vectors, just as the family of Laughlin states can be identified by 
the shape (and orientation) of their correlation holes, or the metric of their wave functions.

After a sudden quantum quench within the same topological phase, the topology of the system 
is not expected to change. 
However, the dynamical evolution of the system, encoded in the normalized effective pseudo 
magnetic field $\hat B(t, k_1, k_2)$, exhibits complex patterns. 
In general, the evolution of the momentum $k_c(t)$ of the persisting zero modes allows us to 
peek into the dynamical geometry of the topological state, which reveals 
the interband excitations whose weights are dominated by the extrema of the excitation gap. 
Nevertheless, the underlying $\hat B(t, k_c, k_2)$ no longer sweep a great circle on the Bloch sphere. 
Indeed, we can filter away the high-frequency components in $k_2$ for the edge modes 
and obtain a good approximation to $k_c(t)$. 
This approximation effectively defines a plane that intersects the Bloch sphere by a great circle, 
which can be regarded as the counterpart of the pseudo 
magnetic field in the equilibrium case. 
The coplanarity condition for $\hat B(k_c, k_2)$ in the equilibrium case is, thus, 
relaxed to a triple-integral condition for $k_c(t)$, which encrypts the
dynamical geometry through a time-independent metric. 

This naturally allows us to extend the definition of the winding vector $\hat m(k_c)$ 
to the dynamically case and use it to visualize the non-equilibrium topological states. 
Such a definition of a unit vector on the Bloch sphere bears similarities to the winding vector 
in the context of a pair of Dirac points~\cite{Montambaux18,Lim20} 
around which the winding number is defined and 
which rotates during the motion of the Dirac points. 
With the dynamical winding vector represented by points on the Bloch sphere, 
we show that the generic quench dynamics can be quite complex. 
Though the winding vector evolves from that of the pre-quench system to that of the post-quench 
system as expected, the time evolution occurs along a trace that can be significantly larger than 
the distance between the pre- and post-quench equilibrium systems. 
More surprisingly, the dynamical winding vector along the trace may not have an equilibrium 
counterpart even though the pre- and post-quench states lie well inside the topological phase. 
This means that the swirling of the pseudospin magnetic field (see Fig.~\ref{fig:geometry}) 
allows the winding vector, or the perpendicular plane, to tilt to directions not possible 
to reach in the equilibrium case.
Thus, the quantum quench within the same topological phase enlarges the topological phase space, 
even if we are only concerned about the winding vector at the characteristic momentum of the zero modes. 

Strictly speaking, the winding vector trace in the dynamical case is not a faithful representation 
of the non-equilibrium topological states. 
Obviously, different states (even with different pre- and post-quench Hamiltonian) can have the same winding vector. 
The bulk state is also not identical to the corresponding equilibrium state. 
Nevertheless, it allows us to visualize the evolution of the topological states and address questions 
like how close or how fast the system is approaching the post-quench equilibrium state. 
In the case when the chiral symmetry is preserved in both pre- and post-quench Hamiltonian, 
the winding vector is confined to the $y$-$z$ plane by symmetry and 
exhibits oscillations with amplitude decaying in a power law with exponent $-1/2$. 
The oscillation spectrum and the exponent are clear signatures that the dynamical winding vector 
correctly reveal the physics associated with the stationary points of the interband excitation gap.

The same physics behind the power-law evolution of the winding vector is also reflected 
in the evolution of other quantities that can be used to describe the non-equilibrium system, 
although to a less extent. 
For example, when the entanglement crossing momentum $k_c$ is not the same for the pre- 
and post-quench Hamiltonian, its time evolution $k_c(t)$ exhibit a similar oscillations 
with amplitude decaying in a power law with exponent $-1/2$. 
Even though the  oscillation spectrum is different in general, the characteristic frequencies, 
which originate from the same stationary points, are the same.
For the same physics, the entanglement velocity evolves with similar oscillations 
whose amplitude increases in a power law with exponent $1/2$. 
This necessarily causes the velocity to change sign, which leads to the emergence or 
annihilation of pairs of entanglement crossings. 

Therefore, we believe that our formulation and application of the dynamical winding vector
facilitate the study of non-equilibrium properties of topological systems. 
For example, the method can be generalized to systems with higher Chern indices or 
to more complex quench protocols. 
Another interesting direction is to study topological systems with decoherence to explore 
how the geometrical information encoded in the systems is lost to the environment. 

%%%%%%%%%%%%%%%%%%%%
\begin{acknowledgments}
This work was supported by the National Key Research and Development Program of China (2021YFA1401902), the Strategic Priority Research Program of Chinese Academy of Sciences 
through Grant No. XDB28000000, and NSFC project No. 11974308. 
\end{acknowledgments}


\begin{thebibliography}{99}

\bibitem{NielsenChuang}
M. A. Nielsen and I. L. Chuang, 
{\it Quantum Computation and Quantum Information}, 1st ed. 
(Cambridge University Press, Cambridge, England, 2004).

\bibitem{Li08}
H. Li and F. D. M. Haldane, Entanglement spectrum as a generalization of entanglement entropy: Identification of topological order in non-Abelian fractional quantum Hall effect states,
Phys. Rev. Lett. {\bf 101}, 010504 (2008).

% Quantum Hall states
\bibitem{Sterdyniak12}
A. Sterdyniak, A. Chandran, N. Regnault, B. A. Bernevig, and P. Bonderson, Real-space entanglement spectrum of quantum Hall states,
Phys. Rev. B {\bf 85}, 125308 (2012).

\bibitem{Dubail12}
J. Dubail, N. Read, and E. H. Rezayi, Real-space entanglement spectrum of quantum Hall systems,
Phys. Rev. B {\bf 85}, 115321 (2012).

\bibitem{Liu13}
Z. Liu and E. J. Bergholtz, From fractional Chern insulators to Abelian and non-Abelian fractional quantum Hall states: Adiabatic continuity and orbital entanglement spectrum,
Phys. Rev. B {\bf 87}, 035306 (2013).

\bibitem{Rodriguez13}
I. D. Rodriguez, S. C. Davenport, S. H. Simon, and J.  K. Slingerland, Entanglement spectrum of composite fermion states in real space,
Phys. Rev. B {\bf 88}, 155307 (2013).

% Topological band insulators

\bibitem{Fidkowski10}
L. Fidkowski, Entanglement spectrum of topological insulators and superconductors, Phys. Rev. Lett. {\bf 104}, 130502 (2010).

\bibitem{Turner10}
A. M. Turner, Y. Zhang, and A. Vishwanath, Entanglement and inversion symmetry in topological insulators,
Phys. Rev. B {\bf 82}, 241102R (2010).

% Spin chains

\bibitem{Pollmann10}
F. Pollmann, E. Berg, A. M. Turner, and M. Oshikawa, Entanglement spectrum of a topological phase in one dimension
Phys. Rev. B {\bf 81}, 064439 (2010).

\bibitem{Thomale10}
R. Thomale, D. P. Arovas, and B. A. Bernevig, Nonlocal order in gapless systems: Entanglement spectrum in spin chains,
Phys. Rev. Lett. {\bf 105}, 116805 (2010).

% Chern insulators

\bibitem{Prodan10}
E. Prodan, T. L. Hughes, and B. A. Bernevig, Entanglement spectrum of a disordered topological Chern insulator,
Phys. Rev. Lett. {\bf 105}, 115501 (2010).

\bibitem{Huang12a}
Z. Huang and D. P. Arovas, Entanglement spectrum and Wannier center flow of the Hofstadter problem,
Phys. Rev. B {\bf 86}, 245109 (2012).

\bibitem{Huang12b}
Z. Huang and D. P. Arovas, arXiv:1205.6266.

\bibitem{Hermanns14}
M. Hermanns, Y. Salimi, M. Haque, and L. Fritz, Entanglement spectrum and entanglement
Hamiltonian of a Chern insulator with open
boundaries,
J. Stat. Mech. P10030 (2014).

% Kitaev model

\bibitem{Yao10}
H. Yao and X.-L. Qi, Entanglement entropy and entanglement spectrum of the Kitaev model,
Phys. Rev. Lett. {\bf 105}, 080501 (2010).

% Experiment

\bibitem{Choo18}
K. Choo, C. W. von Keyserlingk, N. Regnault, and T. Neupert, Measurement of the entanglement spectrum of a symmetry-protected topological state using the IBM quantum computer,
Phys. Rev. Lett. {\bf 121}, 086808 (2018).

\bibitem{Kokail21}
C. Kokail, R. van Bijnen, A. Elben, B. Vermersch, and P. Zoller, Entanglement Hamiltonian tomography in quantum simulation,
Nat. Phys. {\bf 17}, 936 (2021).

% Bulk-edge

\bibitem{Qi12}
X.-L. Qi, H. Katsura, and A. W. W. Ludwig, General relationship between the entanglement spectrum and the edge state spectrum of topological quantum states,
Phys. Rev. Lett. {\bf 108}, 196402 (2012).

\bibitem{Swingle12}
B. Swingle and T. Senthil, Geometric proof of the equality between entanglement and edge spectra,
Phys. Rev. B {\bf 86}, 045117 (2012).

\bibitem{Chandran11}
A. Chandran, M. Hermanns, N. Regnault, and B. A. Bernevig, Bulk-edge correspondence in entanglement spectra,
Phys. Rev. B {\bf 84}, 205136 (2011).

\bibitem{Cirac11}
J. I. Cirac, D. Poilblanc, N. Schuch, and F. Verstraete, Entanglement spectrum and boundary theories with projected entangled-pair states,
Phys. Rev. B {\bf 83}, 245134 (2011).

%

\bibitem{Peschel03}
I. Peschel, Calculation of reduced density matrices from correlation functions,
J. Phys. A: Math. Gen. {\bf 36}, L205 (2003).

% Quench dynamics

\bibitem{Gong18}
Z. Gong and M. Ueda, Topological entanglement-spectrum crossing in quench dynamics,
Phys. Rev. Lett. {\bf 121}, 250601 (2018).

\bibitem{Chang18}
P.-Y. Chang, Topology and entanglement in quench dynamics,
Phys. Rev. B {\bf 97}, 224304 (2018).

\bibitem{Lu19}
S. Lu and J. Yu, Stability of entanglement-spectrum crossing in quench dynamics of one-dimensional gapped free-fermion systems,
Phys. Rev. A {\bf 99}, 033621 (2019).

% non-hermitian

\bibitem{Sayyad21}
S. Sayyad, J. Yu, A. G. Grushin, and L. M. Sieberer, Entanglement spectrum crossings reveal non-Hermitian dynamical topology,
Phys. Rev. Research {\bf 3}, 033022 (2021).

\bibitem{Hermanns22}
C. Ortega-Taberner, L. R{\o}dland, and M. Hermanns, Polarization and entanglement spectrum in non-Hermitian systems,
Phys. Rev. B {\bf 105}, 075103 (2022).

%dynamics, topology



\bibitem{Unal16} F. Nur \"{U}nal, E. J. Mueller, and M. \"{O}. Oktel, Nonequilibrium fractional Hall response after a topological quench, Phys. Rev. A {\bf 94}, 053604 (2016).

\bibitem{Wang17} C. Wang, P. Zhang, X. Chen, J. Yu, and H. Zhai, Scheme to measure the topological number of a Chern Insulator from quench dynamics, Phys. Rev. Lett. {\bf 118}, 185701 (2017).

\bibitem{Sun18} W. Sun \textit{et al}, Uncover topology by quantum quench dynamics, Phys. Rev. Lett. {\bf 121}, 250403 (2018).

\bibitem{Tarnowski19} M. Tarnowski, F. Nur \"{U}nal, N. Fl\"{a}schner, B. S. Rem, A. Eckardt, K. Sengstock, and C. Weitenberg, Measuring topology from dynamics by obtaining the Chern number from a linking number, Nat. Commun. {\bf 10}, 1728 (2019).

% FQHE

\bibitem{Haldane11}
F. D. M. Haldane, Geometrical description of the fractional quantum Hall effect, Phys. Rev. Lett. {\bf 107}, 116801 (2011).

\bibitem{Qiu12}
R.-Z. Qiu, F. D. M. Haldane, X. Wan, K. Yang, and S. Yi, Model anisotropic quantum Hall states, Phys. Rev. B {\bf 85}, 115308 (2012).

\bibitem{Liu18}
Z. Liu, A. Gromov, and Z. Papi\'c, Geometric quench and nonequilibrium dynamics of fractional quantum Hall states,
Phys. Rev. B {\bf 98}, 155140 (2018).

\bibitem{Liu21}
Z. Liu, A. C. Balram, Z. Papi\'c, and A. Gromov, Quench dynamics of collective modes in fractional quantum Hall bilayers,
Phys. Rev. Lett. {\bf 126}, 076604 (2021).

\bibitem{Ji22}
H.-X. Ji, L.-H. Mo, and X. Wan, Dynamics of the entanglement zero modes in the Haldane model under a quantum quench,
Chin. Phys. Lett. {\bf 39}, 030301 (2022).

\bibitem{Montambaux18}
G. Montambaux, L.-K. Lim, J.-N. Fuchs, and F. Pi\'echon, Winding vector: How to annihilate two Dirac points with the same charge,
Phys. Rev. Lett. {\bf 121}, 256402 (2018).

\bibitem{Lim20}
L.-K. Lim, J.-N. Fuchs, F. Pi\'echon, and G. Montambaux, Dirac points emerging from flat bands in Lieb-kagome lattices,
Phys. Rev. B {\bf 101}, 045131 (2020).

\bibitem{Haldane88}
F. D.M. Haldane, Model for a quantum Hall effect without Landau levels: Condensed-matter realization of the "parity anomaly", Phys. Rev. Lett. {\bf 61}, 2015 (1988).

\bibitem{Chung01}
M.-C. Chung and I. Peschel, Density-matrix spectra of solvable fermionic systems, Phys. Rev. B {\bf 64}, 064412 (2001).

\bibitem{Delplace11}
P. Delplace, D. Ullmo, and G. Montambaux, Zak phase and the existence of edge states in graphene,
Phys. Rev. B {\bf 84}, 195452 (2011).

\bibitem{Calabrese05}
P. Calabrese and J. Cardy, Evolution of entanglement entropy in one-dimensional systems, J. Stat. Mech. P04010 (2005).

\bibitem{Calabrese06}
P. Calabrese and J. Cardy, Time dependence of correlation functions following a quantum quench, Phys. Rev. Lett. {\bf 96}, 136801 (2006).

\bibitem{Makki22}
A. A. Makki, S. Bandyopadhyay, S. Maity, and A. Dutta, Dynamical crossover behavior in the relaxation of quenched quantum many-body systems, Phys. Rev. B {\bf 105}, 054301 (2022). 

\bibitem{Sim22} K. Sim, R. Chitra, and P. Molignini, Quench dynamics and scaling laws in topological nodal loop semimetals, Phys. Rev. B {\bf 106}, 224302 (2022). 

\bibitem{Ramos23}
F. B. Ramos, A. Urichuk, I. Schneider, and J. Sirker, Power-law decay of correlations after a global quench in the massive XXZ chain, Phys. Rev. B {\bf 107}, 075138 (2023). 

\bibitem{Cao24} K. Cao, Y. Hu, P. Tong, and G. Yang, Dynamical relaxation behavior of an extended XY chain with a gapless phase following a quantum quench, Phys. Rev. B {\bf 109}, 024303 (2024).

\bibitem{Lazarides14} A. Lazarides, A. Das, and R. Moessner, Equilibrium states of generic quantum systems subject to periodic driving, Phys. Rev. E {\bf 90}, 012110 (2014).

\bibitem{Kiendl17} T. Kiendl and F. Marquardt, Many-particle dephasing after a quench, Phys. Rev. Lett. {\bf 118}, 130601 (2017).

\bibitem{Kormos17} M. Kormos, M. Collura, G. Tak\'{a}cs, and P. Calabrese, Real-time confinement following a quantum quench to a non-integrable model, Nat. Phys. {\bf 13}, 246 (2017).

\bibitem{CAlvaredo20} O. A. Castro-Alvaredo, M. Lencs\'{e}s, I. M. Sz\'{e}cs\'{e}nyi, and J. Viti, Entanglement oscillations near a quantum critical point, Phys. Rev. Lett. {\bf 124}, 230601 (2020).

\bibitem{Serbyn21} M. Serbyn, D. A. Abanin and Z. Papi\'{c}, Quantum many-body scars and weak breaking of ergodicity, Nat. Phys. {\bf 17}, 675 (2021).

\bibitem{Delfino22} G. Delfino and M. Sorba, Persistent oscillations after quantum quenches in d dimensions, Nucl. Phys. B {\bf 974}, 115643 (2022).

\bibitem{Robertson24} J. H. Robertson, R. Senese, and F. H. L. Essler, Decay of long-lived oscillations after quantum quenches in gapped interacting quantum systems, Phys. Rev. A {\bf 109}, 032208 (2024).


\end{thebibliography}
\end{document}